\documentclass[12pt]{article}

\usepackage[dvipdfm]{graphicx}
\usepackage{booktabs}
\usepackage{bm}
\usepackage{amsmath,amssymb}
\allowdisplaybreaks
\usepackage{cite}

\makeatletter
    
    \@addtoreset{equation}{section}
\makeatother

\newcommand{\fm}{\text{fm}}
\newcommand{\MeV}{\text{MeV}}
\newcommand{\VEV}[1]{\left\langle #1\right\rangle}

\newcommand{\cl}{\text{cl}}
\newcommand{\drv}[2]{\frac{d #1}{d #2}}
\newcommand{\EQ}{\text{EQ}}

\newcommand{\p}{\partial}
\newcommand{\fpi}{f_\pi}
\newcommand{\mpi}{m_\pi}
\newcommand{\bmx}{\bm{x}}
\newcommand{\bmy}{\bm{y}}
\newcommand{\abs}[1]{\left| #1\right|}

\newcommand{\tr}{\mathop{\rm tr}}
\newcommand{\Tr}{\mathop{\rm Tr}}
\newcommand{\Drv}[2]{\frac{\p #1}{\p #2}}

\newcommand{\wt}[1]{\widetilde{#1}}
\newcommand{\wh}[1]{\widehat{#1}}

\newcommand{\calO}{{\mathcal O}}
\newcommand{\calL}{{\mathcal L}}
\newcommand{\calI}{{\mathcal I}}
\newcommand{\calJ}{{\mathcal J}}
\newcommand{\calW}{{\mathcal W}}

\newcommand{\Lcl}{L^\cl}
\newcommand{\Rcl}{R^\cl}
\newcommand{\calLcl}{\calL^\cl}
\newcommand{\Ucl}{U_\cl}
\newcommand{\Mcl}{M_\cl}
\newcommand{\RiRd}{R^{-1}\dot{R}}

\newcommand{\zbmy}{\wt{\bm{y}}}
\newcommand{\zy}{\wt{y}}

\newcommand{\Pmatrix}[1]{\begin{pmatrix} #1\end{pmatrix}}

\newcommand{\dtRiRd}{\calW}

\newcommand{\why}{\wh{y}}

\newcommand{\yx}[1]{\Upsilon_{#1}}
\newcommand{\yxx}[2]{\Gamma^{#1}_{#2}}

\newcommand{\bmOmg}{\bm{\Omega}}
\newcommand{\QLp}{\text{QL-part}}

\newcommand{\Mcldl}{\wh{M}_\cl}
\newcommand{\calIdl}{\wh{\calI}}
\newcommand{\calJdl}{\wh{\calJ}}
\newcommand{\bmOmgdl}{\wh{\bm{\Omega}}}
\newcommand{\rdl}{\rho}

\newcommand{\dlmbd}{\delta_\lambda}
\newcommand{\dtt}{\delta_\varepsilon}

\newcommand{\fpiesq}{\fpi^2 e^2}

\newcommand{\bra}[1]{\left\langle #1\right\vert}
\newcommand{\ket}[1]{\left\vert #1\right\rangle}

\newcommand{\JBclz}{J_B^{\cl\;0}}

\newcommand{\veps}{\varepsilon}
\newcommand{\smallPmatrix}[1]{
   \left(\begin{smallmatrix}#1 \end{smallmatrix}\right)}

\begin{document}
\baselineskip=16.8pt plus 0.2pt minus 0.1pt

\begin{titlepage}

\title{
\hfill\parbox{4cm}{\normalsize KUNS-2289}\\[1cm]
{\Large\bf
Relativistic Collective Coordinate System of Solitons
and Spinning Skyrmion}
}

\author{
Hiroyuki {\sc Hata}\footnote{
{\tt hata@gauge.scphys.kyoto-u.ac.jp}}
\ and
Toru {\sc Kikuchi}\footnote{
{\tt kikuchi@gauge.scphys.kyoto-u.ac.jp}}
\\[7mm]
{\it
Department of Physics, Kyoto University, Kyoto 606-8502, Japan
}
}

\date{{\normalsize August 2010}}
\maketitle

\begin{abstract}
\normalsize
We consider constructing the relativistic system of collective
coordinates of a field theory soliton on the basis of a simple
principle: The collective coordinates must be introduced into the
static solution in such a way that the equation of motion of the
collective coordinates ensures that of the original field theory.
As an illustration, we apply this principle to the quantization of
spinning motion of the Skyrmion by incorporating the leading
relativistic correction to the rigid body approximation.
We calculate the decay constant and various static properties of
nucleons, and find that the relativistic corrections are in the range
of $5\%$ -- $20\%$.
We also examine how the baryons deform due to the spinning motion.
\end{abstract}

\thispagestyle{empty}
\end{titlepage}

\section{Introduction}
\label{sec:Intro}

Collective coordinates of a field theory soliton play a central role
in the study of the soliton dynamics (see, e.g.,
\cite{Rajaraman} for a review).
They represent the motion of the soliton as a whole in the symmetry
directions of the original field theory.
Since the soliton breaks these symmetries, the collective coordinates
are related with the zero-mode fluctuations around the soliton.
The simplest way to construct the system of collective coordinates is
to promote the (originally constant) free parameters of the static
soliton associated with symmetries to time-dependent dynamical
variables by discarding the fluctuation of non-zero modes.

As an example, let us consider the collective coordinate of
space-translation of a static soliton $\phi_\cl(x)$ in a scalar field
theory in $1+1$ dimensions with Lagrangian density
$\calL=-(1/2)(\p_\mu\phi)^2-U(\phi)$.
In the simple treatment, we take
\begin{equation}
\phi(x,t)=\phi_\cl\bigl(x-X(t)\bigr) ,
\label{phicl(x-X)}
\end{equation}
as the soliton field with a dynamical variable of the center-of-mass
$X(t)$. Plugging this into the Lagrangian density and carrying out the
space integration, we get the following Lagrangian of $X(t)$:
\begin{equation}
L=-m +\frac12 m\dot{X}(t)^2 ,
\end{equation}
with $m$ being the energy of the static soliton.
Somehow, we have obtained the {\em non-relativistic} Lagrangian of
a point particle with coordinate $X(t)$, although we started with a
relativistic scalar field theory.
We need an improved treatment of collective coordinates which leads to
the relativistic Lagrangian
\begin{equation}
L=-m\sqrt{1-\dot{X}^2}\, .
\label{L_X_R}
\end{equation}
The purpose of this paper is to propose a simple and general principle
of constructing the system of collective coordinates of a field theory
soliton, by which we can obtain the complete and relativistic
Lagrangian of collective coordinates.

First, let us state our principle of introducing the collective
coordinates into a static soliton solution: {\em
Collective coordinates must be introduced into the static solution in
such a way that the equation of motion (EOM) of the collective
coordinates ensures that of the original field theory.}
We call this the ``EOM principle'' hereafter.
Although the original field theory has an infinite number of degrees
of freedom, the reduced system of collective coordinates has only a
finite number of ones. Therefore, even if the EOM of collective
coordinates holds, it does not necessarily imply that the field theory
EOM holds. Our EOM principle demands in a sense that there should be
no mismatch between the field theory dynamics and the collective
coordinate dynamics. If the original field theory is a relativistic
one, our EOM principle should automatically lead to a relativistic
Lagrangian of the collective coordinates.

Practically, the introduction of collective coordinates by the EOM
principle must be carried out in a self-consistent manner.
Starting with a suitably chosen dependence of the soliton field on the
collective coordinates, we obtain the Lagrangian and consequently the
EOM of the collective coordinates by plugging the soliton field into
the field theory Lagrangian.
Besides this, we plug the assumed soliton field into the field theory
EOM, and examine whether it holds under the EOM of collective
coordinates.
We tune the dependence of the soliton field on the collective
coordinates so that the EOM principle is realized.

In the above example of the center-of-mass in a scalar field theory in
$1+1$ dimensions, the field theory EOM is in fact broken by the term
$\dot{X}^2\phi''_\cl(x-X)$ if we take \eqref{phicl(x-X)} as the
soliton field (see eq.\ \eqref{EOM_X}).
Let us apply our EOM principle to this system to resolve the breaking
of field theory EOM successively in the number of time-derivatives.
As an improved soliton field, we take
$\phi(x,t)=\phi_\cl\bigl(y(x,t)\bigr)$ with
\begin{equation}
y(x,t)=\bigl(x-X(t)\bigr)\left(1+c_2\dot{X}(t)^2+c_4\dot{X}(t)^4
+\cdots\right) ,
\label{y(x,t)}
\end{equation}
where $c_2$ and $c_4$ are constants to be determined by the EOM
principle. The corresponding Lagrangian of $X(t)$ reads
\begin{equation}
L=\int\!dx\,\calL\bigr|_{\phi(x,t)=\phi_\cl(y(x,t))}
=m\left\{-1+\frac12\dot{X}^2
+\frac12 c_2(1-c_2)\dot{X}^4+\cdots\right\} ,
\label{L_X}
\end{equation}
up to time-derivative terms.
Therefore, the EOM of $X(t)$ remains unchanged from that in the simple
treatment: $\ddot{X}=0$.
On the other hand, for the field theory EOM, we have
\begin{equation}
-\p_\mu^2\phi(x,t)+U'\bigl(\phi(x,t)\bigr)=\left[
(1-2 c_2)\dot{X}^2 +\left(2c_2-c_2^2-2 c_4\right)\dot{X}^4
+\cdots\right]\phi''_\cl(y) ,
\label{EOM_X}
\end{equation}
up to terms containing $\ddot{X}$.
{}From this, we find that the EOM principle is satisfied (up to
$\dot{X}^6$) by taking $c_2=1/2$ and $c_4=3/8$.
Then, $y(x,t)$ \eqref{y(x,t)} and $L$ \eqref{L_X} are nothing but the
first three terms of the expansion in powers of $\dot{X}^2$ of the
Lorentz boost
\begin{equation}
y(x,t)=\frac{x-X(t)}{\sqrt{1-\dot{X}^2}} ,
\label{boost}
\end{equation}
and the relativistic Lagrangian \eqref{L_X_R},
respectively.\footnote{
For a constant $\dot{X}$, $y(x,t)$ of \eqref{boost} obviously leads to
the relativistic Lagrangian \eqref{L_X_R} to all orders in
$\dot{X}^2$. However, for a generic $X(t)$, we have to add
\eqref{boost} corrections of $O(\p_t^6)$ starting with
$\frac23(x-X(t))^3(\dot{X}\ddot{X})^2$ for obtaining \eqref{L_X_R}.
}

Here, we should mention another and conventional way of constructing
the system of collective coordinates. It is, starting with suitably
introduced collective coordinates (for example, \eqref{phicl(x-X)} for
the center-of-mass), to integrate over the non-zero modes around the
soliton. Equivalently, we solve the EOM of the non-zero modes to
express them in terms of the collective coordinates.
Since the set of EOMs of both the zero and non-zero modes are
equivalent to the field theory EOM, integration over the non-zero
modes leads to a system of collective coordinates whose EOM implies
the field theory EOM.
In fact, the relativistic energy of the center-of-mass
motion, $E=\sqrt{P^2+m^2}$, has been obtained by this method
for a scalar field theory in $1+1$ dimensions \cite{GS,GJS,GJS2}.
Looking back the procedure of \eqref{y(x,t)}, \eqref{L_X} and
\eqref{EOM_X}, the EOM of the non-zero modes implies that they are
equal to zero if we adopt the elaborated way of introducing the
center-of-mass coordinate $X(t)$ by using $y(x,t)$ of \eqref{y(x,t)}.

In the case of the center-of-mass coordinate of a soliton, we know its
relativistic Lagrangian \eqref{L_X_R} from the start.
However, for the collective coordinate of the spinning motion of a
soliton, it is quite a non-trivial problem to obtain its relativistic
description, and this is our main concern of this paper.
Concretely, we consider the spinning collective coordinate of the
Skyrmion, which is a soliton representing a baryon in the low energy
effective theory of pions (the Skyrme-model) \cite{SkyrmeI}.
Let $\Ucl(\bmx)$ be the static Skyrmion solution.
In the simple treatment, we take
$U(\bmx,t)=\Ucl\bigl(R^{-1}(t)\bmx\bigr)$ as
the spinning Skyrmion field with the rotation matrix $R(t)$ of
spinning motion.
This leads to the Hamiltonian of a spinning spherical rigid body
\cite{ANW,AN}:
\begin{equation}
H_\text{rigid body}=\Mcl +\frac12 \calI\bmOmg^2
=\Mcl +\frac{1}{2\calI}\bm{I}^2 ,
\label{H_rigidbody}
\end{equation}
where $\Mcl$, $\calI$, $\bmOmg$ and $\bm{I}(=\calI\bmOmg)$
are the energy of the static solution, moment of inertia, angular
velocity in isospace (body-fixed frame) and the isospin operator,
respectively.
($H_\text{rigid body}$ also has an equivalent expression with
$\bmOmg$ and $\bm{I}$ replaced with the angular velocity $\bm{\omega}$
in real space and the spin operator $\bm{J}(=\calI\bm{\omega})$,
respectively.)
However, this rigid body description is certainly a non-relativistic
one valid only for a slow spinning motion.
For a large angular velocity, the Skyrmion should intuitively deform
from its original spherical shape, and this should lead to corrections
with higher powers of $\bmOmg$ to the rigid body Hamiltonian.
Therefore, it is an interesting theoretical subject to obtain
relativistic (i.e., higher time-derivative) corrections to the rigid
body approximation \eqref{H_rigidbody}, or, if possible, to find the
fully relativistic description which is the spinning motion counterpart
of \eqref{L_X_R}.

Relativistic corrections to the rigid body approximation
\eqref{H_rigidbody} may be important also phenomenologically for the
Skyrmion describing baryons.
Identifying the eigenvalues of the Hamiltonian \eqref{H_rigidbody}
with the masses of the nucleon ($M_N=939\,\MeV$,
$\bm{I}^2=\bm{J}^2=3/4$) and $\Delta$ ($M_\Delta=1232\,\MeV$,
$\bm{I}^2=\bm{J}^2=15/4$), we find the followings:
\begin{equation}
\Mcl=866\,\MeV,\quad \calI=0.00512\,\MeV^{-1},\quad
\abs{\bmOmg_N}=169\,\MeV,\quad
\abs{\bmOmg_\Delta}=378\,\MeV .
\label{valuesRB}
\end{equation}
This implies that, (i) about $8\%$ ($30\%$) of the total mass of
the nucleon ($\Delta$) comes from the rotational energy, and (ii)
the rotational velocity of the nucleon ($\Delta$) at the radius of
$1\,\fm$ is $86\%$ ($190\%$ !) of the light velocity.
These facts suggest that relativistic corrections are non-negligible
especially for $\Delta$, and that we should reexamine the analysis of
Refs.\ \cite{ANW,AN} based on the rigid body approximation by
incorporating the effects of the corrections.
The limitations of the rigid body approximation were shown numerically
without relying on the expansion in powers of angular velocity
in the case of the $(2+1)$-dimensional baby version of the Skyrme
theory \cite{Piette:1994mh}.

In this paper, we derive the leading relativistic correction to the
rigid body Lagrangian of the spinning Skyrmion by taking the EOM
principle as the basic principle of constructing the Lagrangian of
spinning collective coordinate.
We follow the procedure starting with \eqref{y(x,t)} for the
center-of-mass motion. Namely, assuming the spinning Skyrmion field of
the form $U(\bmx,t)=\Ucl\bigl(\bmy(\bmx,t)\bigr)$ with
$\bmy(\bmx,t)=R(t)^{-1}\bmx +\mbox{(relativistic correction term)}$,
we determine the correction term from the EOM principle.
Note that the relativistic correction to $U(\bmx,t)$ is made only to
the argument of the static solution.
In the case of the center-of-mass motion, the first relativistic
correction is specified by the constant coefficient $c_2$.
For the spinning Skyrmion, we have to determine three
functions $(A(r),B(r),C(r))$ of the radial coordinate $r$ appearing in
$\bmy(\bmx,t)$ (see eq.\ \eqref{bmy=}) by solving their differential
equations derived from the EOM principle.
Then, from $(A(r),B(r),C(r))$, the $\bmOmg^4$ correction term to the
rigid body Lagrangian is obtained (see eq.\ \eqref{L_R}).
{}From the function $\bmy(\bmx,t)$, we can also learn how the static
Skyrmion of spherical shape deforms due to the spinning motion.

Then, we repeat the analysis of Refs.\ \cite{ANW,AN} for the decay
constant $\fpi$ and various static properties of nucleons, such as
charge radii, magnetic moments and axial vector coupling, by
including the relativistic corrections.
We find that the contribution of the relativistic corrections is in
the range of $5\%$ to $20\%$. However, the comparison with the
experimental data is a rather disappointing one:
Although the value of $\fpi$ is shifted closer to the experimental one
due to the relativistic correction, it makes the theoretical
value move away from the experimental one for most of the static
properties.
However, this should not be regarded as a problem of our EOM
principle. Either higher order relativistic corrections are important,
or we cannot expect the Skyrme model to reproduce the baryon sector so
precisely.

As we mentioned before, there is another way of constructing the
system of collective coordinates. This is, starting with a simply
introduced collective coordinates (such as in the rigid body
approximation), to solve the EOM of the massive modes to express them
in terms of the collective coordinates. Though there has appeared no
explicit analysis of the spinning Skyrmion by this method, the present
one using the EOM principle should have some advantages: Firstly, ours
is simpler since we are not bothered by the non-zero modes, and
secondly we can directly know how the soliton deforms due to its fast
spinning motion since the collective coordinates are introduced by
deforming the coordinate of the static solution.
We should mention that there have appeared many attempts
to improve the rigid body approximation of \cite{ANW,AN}.
They include the papers
\cite{Bander:1984gr,Hajduk:1984as,Hayashi:1984rz,
Liu:1984rm,Braaten:1984qe,Rajaraman:1985ty,Hajduk:1986zw,
Wambach:1986ut,Li:1987kq,Braaten:1988cc,Schroers:1993yk,
Schwesinger:1993ia,Thomas:1993bx,Dorey:1994fk,Rakhimov:1996qe,
Yakhshiev:2001ht,Yakhshiev:2002xf,Battye:2005nx,Houghton:2005iu,
Cherman:2005hy,Fortier:2008yj}.
We wish to emphasize that our method based on the EOM principle gives
a systematic and self-consistent way of analyzing the spinning motion
of a field theory soliton without any ad hoc physical pictures.

Finally, we remark that the EOM principle has already been used
for solitons in gauge theories even in the ``rigid body
approximation'' before introducing the relativistic corrections.
Namely, the dependence of the gauge field on the collective
coordinates must be determined from the EOM of $A_0$, the Gauss law,
which is non-trivial even in the rigid body approximation.
The examples are the magnetic monopoles in gauge-Higgs systems
(see \cite{Harvey} for a review) and the baryon solution
\cite{HSSY,HM} in the Sakai-Sugimoto model \cite{SS,SS2}.

The organization of the rest of this paper is as follows.
In Sec.\ \ref{sec:RB}, we summarize the rigid body approximation to
the spinning Skyrmion and show how our EOM principle is violated.
In Sec.\ \ref{sec:realizeEOMP}, we introduce the improved spinning
Skyrmion field with the leading relativistic correction to obtain, on
the basis of the EOM principle, the differential equations and the
boundary conditions of the functions $(A(r),B(r),C(r))$ specifying the
correction. We also consider the deformation of the spinning Skyrmion
caused by the relativistic correction.
The Lagrangian of the spinning Skyrmion with the leading relativistic
correction is given in Sec.\ \ref{sec:RLagrangian}.
In Sec.\ \ref{sec:detfpi}, we determine the values of the parameters
$(\fpi,e)$ of the Skyrme theory from the masses of the nucleon,
$\Delta$ and the pion.
The profiles of the functions and the integrands related with the
corrections are given. The deformations of the baryons are also
pictorially shown there.
Then, we present the analysis of the static properties of nucleons.
The final section (Sec.\ \ref{sec:SD}) is devoted to a summary and
future problems.
In the Appendices (A--E), we present technical details and complicated
equations used in the text.

The present paper is a detailed version of \cite{HK}, where the EOM
principle was first proposed and applied to the quantization of the
spinning motion of the Skyrmion.

\section{Skyrmion in the rigid body approximation
and its limitation}
\label{sec:RB}

In this section, we summarize the Skyrmion and the quantization of its
spinning degrees of freedom in the rigid body approximation
\cite{ANW,AN}. We also explain how the EOM principle is violated in
this approximation.
This is the starting point of our relativistic
extension. Another purpose of this section is to fix our notations and
conventions.

We consider the $SU(2)$ Skyrme model \cite{SkyrmeI} described by the
chiral Lagrangian with the Skyrme term:
\begin{equation}
\calL=\tr\left\{-\frac{\fpi^2}{16}L_\mu^2
+\frac{1}{32e^2}[L_\mu,L_\nu]^2
+\frac{\fpi^2}{8}\mpi^2(U-\bm{1}_2)\right\} ,
\label{calL_Skyrme}
\end{equation}
where $U(x)$ is an $SU(2)$-valued scalar field, $L_\mu$ is defined by
\begin{equation}
L_\mu=-iU\p_\mu U^\dagger ,
\label{L_mu}
\end{equation}
and $e$ is a dimensionless parameter. Our flat space-time metric is
$\eta_{\mu\nu}=\text{diag}\left(-1,1,1,1\right)$, though not written
explicitly in \eqref{calL_Skyrme}.
As we will see later in Sec.\ \ref{sec:RLagrangian}, the inclusion of
non-zero pion mass $\mpi$ is indispensable for our relativistic
corrections to be finite.
The EOM of the Lagrangian \eqref{calL_Skyrme} reads
\begin{equation}
\p_\mu\left(L^\mu-\frac{1}{\fpiesq}
\bigl[L_\nu,\bigl[L^\mu,L^\nu\bigr]\bigr]\right)
-i\mpi^2\left(U-\frac12\tr U\right)=0.
\label{FEOM}
\end{equation}
The Skyrme model has a topologically conserved current $J_B^\mu$:
\begin{equation}
J_B^\mu=-\frac{i}{24\pi^2}\epsilon^{\mu\nu\rho\sigma}
\tr(L_\nu L_\rho L_\sigma) ,
\label{JB}
\end{equation}
with $\epsilon^{0123}=1$. The conserved charge is identified with the
baryon number $N_B$:
\begin{equation}
N_B=\int\! d^3x J_B^0 .
\label{eq:N_B}
\end{equation}
Next, the diagonal $SU(2)$ symmetry of the theory gives the conserved
isospin current:
\begin{equation}
J_{V,a}^\mu=\tr\left[\left(J_L^\mu + J_R^\mu\right)\tau_a\right],
\label{JV}
\end{equation}
where the left and right chiral currents $J_{L/R}^\mu$ are defined by
\begin{equation}
J_L^\mu=-\frac{\fpi^2}{16}\left(
L^\mu-\frac{1}{\fpiesq}[L_\nu,[L^\mu,L^\nu]]\right),
\qquad
J_R^\mu=J_L^\mu\bigr|_{L_\mu\text{ replaced with }R_\mu} ,
\label{eq:J_LJ_R}
\end{equation}
with $R_\mu= -iU^\dagger\p_\mu U$.
On the other hand, the conservation of the axial $SU(2)$ current,
\begin{equation}
J_{A,a}^\mu=\tr\left[\left(J_L^\mu - J_R^\mu\right)\tau_a\right],
\label{JA}
\end{equation}
is broken due to the pion mass term;
$\p_\mu J_{A,a}^\mu=-i(\fpi^2\mpi^2/8)\,\tr\left(U\tau_a\right)$.

The Skyrme model has static soliton solutions which are called
Skyrmion and represent the baryons \cite{SkyrmeI}.
The simplest one takes the hedgehog form:
\begin{equation}
\Ucl(\bmx)=\exp\bigl(i\wh{\bmx}\cdot\bm{\tau}F(r)\bigr)
=\cos F(r) +i \wh{\bmx}\cdot\bm{\tau}\sin F(r) ,
\label{hedgehog}
\end{equation}
with $r=\abs{\bmx}$ and $\wh{\bmx}={\bmx}/\abs{\bmx}$.
The function $F(r)$ is subject to the following differential equation:
\begin{equation}
\drv{^2 F}{r^2}+\frac{2}{r}\drv{F}{r}-\frac{\sin 2F}{r^2}
+\frac{4}{\fpiesq}\!
\left[\frac{2\sin^2 F}{r^2}\drv{^2 F}{r^2}
+\frac{\sin 2F}{r^2}\left(\drv{F}{r}\right)^2
-\frac{\sin^2 F\sin 2F}{r^4}\right]\!
-\mpi^2\sin F=0 ,
\label{DEF}
\end{equation}
which is obtained by substituting \eqref{hedgehog} into the EOM
\eqref{FEOM}, or equivalently, by minimizing the mass (energy) $\Mcl$
of the configuration \eqref{hedgehog}:
\begin{equation}
\Mcl=\frac{\pi\fpi^2}{2}\int_0^\infty\!dr\,r^2\left\{(F')^2
+\frac{2\sin^2 F}{r^2}+\frac{4}{\fpiesq}\frac{\sin^2 F}{r^2}
\left[2(F')^2+\frac{\sin^2 F}{r^2}\right]
+2\mpi^2(1-\cos F)\right\} ,
\label{Mcl}
\end{equation}
with the prime on $F$ denoting a differentiation with respect to $r$.
For the hedgehog \eqref{hedgehog} to be non-singular both at the
origin and the infinity and for the mass \eqref{Mcl} to be finite, we
must have $F(0)=n\pi$ and $F(\infty)=0$ (mod $2\pi$) with the integer
$n$ being equal to the baryon number $N_B$.
Restricting ourselves to the solution with unit baryon number $N_B=1$,
the behavior of $F(r)$ near the origin is given by
\begin{equation}
F(r)=\pi - \kappa r +O(r^3)\qquad (r\to 0),
\label{F_origin}
\end{equation}
while that near the infinity reads
\begin{equation}
F(r)=\frac{a}{r}\left(1+\frac{1}{\mpi r}\right)
e^{-\mpi r} +O(e^{-2\mpi r})\qquad (r\to\infty),
\label{F_infty}
\end{equation}
with $\kappa$ and $a$ being constants.
In particular, the power part multiplying $e^{-\mpi r}$ in
\eqref{F_infty} is exact.
The hedgehog solution \eqref{hedgehog} specified by a single function
$F(r)$ represents a spherically symmetric extended object;
the energy density depends only on $r$.

The hedgehog solution \eqref{hedgehog} has two kinds of collective
coordinates; the center-of-mass motion and the space (isospace)
rotation. Note that, for the hedgehog \eqref{hedgehog}, the space
rotation is equivalent to the isospace rotation.
Namely, for any orthogonal matrix $\calO$, we have
\begin{equation}
\Ucl(\calO^{-1}\bmx)=W \Ucl(\bmx)W^{-1} ,
\label{hedgehogP}
\end{equation}
where $W$ is the $SU(2)$ matrix corresponding to $\calO$;
$\calO_{ab}\tau_b=W\tau_a W^{-1}$.
In the quantization of the spinning collective coordinate in the rigid
body approximation \cite{ANW,AN}, we assume that the spinning Skyrmion
field $U(\bmx,t)$ is simply given by
\begin{equation}
U(\bmx,t)=\Ucl\bigl(R^{-1}(t)\bmx\bigr) ,
\label{U_RB}
\end{equation}
where the orthogonal matrix $R(t)$ is the dynamical variable to be
quantized. Since the classical solution $\Ucl(\bmx)$ is of spherical
shape, so does the Skyrme field of \eqref{U_RB}.
An important property of $U(\bmx,t)$ \eqref{U_RB} is
that the left and the right $SO(3)$ transformations on the matrix
$R(t)$ correspond to the rotations in the real space and the
isospace, respectively:
\begin{equation}
U(\bmx,t)\bigr|_{R\to \calO^{-1}_\text{real}R\calO_\text{iso}}
=W_\text{iso}U\bigl(\calO_\text{real}\bmx,t\bigr)W^{-1}_\text{iso}.
\label{ImpProp}
\end{equation}
Therefore, we write the components of $R(t)$ by $R_{ia}(t)$, with
$i$ and $a$ being the real and the isospace indices, respectively.

The dynamics of $R(t)$ is governed by its Lagrangian obtained by
inserting \eqref{U_RB} into the Skyrme model Lagrangian density
\eqref{calL_Skyrme} and carrying out the space integration. It turns
out to be the Lagrangian of a spherical rigid rotor:
\begin{equation}
L_\text{rigid body}(R,\dot{R})
=\int\! d^3x\,\calL\bigr|_{U(\bmx,t)=\Ucl(R^{-1}\bmx)}
=-\Mcl + \frac12\calI\bmOmg^2 ,
\label{L(R,dotR)_RB}
\end{equation}
where
\begin{equation}
\calI=\frac{2\pi\fpi^2}{3}\int_0^\infty\! dr\,r^2\sin^2 F
\left\{1+\frac{4}{\fpiesq}\left[(F')^2+\frac{\sin^2 F}{r^2}
\right]\right\} ,
\label{calI}
\end{equation}
is the moment of inertia, and
\begin{equation}
\Omega_a=\frac12\epsilon_{abc}(R^{-1}\dot R)_{bc} ,
\label{Omega_a}
\end{equation}
is the angular velocity in isospace (which is equal to the angular
velocity in the body-fixed frame).
The angular velocity $\omega_i$ in real space is given by
\begin{equation}
\omega_i=-\frac12\epsilon_{ijk}(\dot{R}R^{-1})_{jk}
=\Omega_i\bigr|_{R\to R^{-1}}=-R_{ia}\Omega_a
\label{omega_i}
\end{equation}
Note that we have
\begin{equation}
\bmOmg^2=\bm{\omega}^2=-\frac12\Tr(\RiRd)^2 .
\label{bmOmg^2=bmomg^2=}
\end{equation}

The Lagrangian \eqref{L(R,dotR)_RB} describes baryons with unit baryon
number $N_B=1$ and zero center-of-mass momentum. They are
specified by the quantum numbers of the spin $\bm{J}=\calI\bm{\omega}$
and the isospin $\bm{I}=\calI\bmOmg$, which are the Noether charges
corresponding to the symmetry transformations of the Lagrangian
\eqref{L(R,dotR)_RB}:
\begin{equation}
R(t)\to \calO^{-1}_\text{real}R(t)\calO_\text{iso} .
\label{RtoOinvRO}
\end{equation}
Since $\bm{I}$ and $\bm{J}$ are related by
$J_i=I_i\bigr|_{R\to R^{-1}}=-R_{ia}I_a$ in the present system, we
have $\bm{I}^2=\bm{J}^2$, which is consistent with the real baryon
spectrum. The decay constant $\fpi$ and various static properties of
nucleons including charge radii, magnetic moments and axial vector
coupling have been calculated in \cite{ANW} ($\mpi=0$ case) and
\cite{AN} by using the masses of the nucleon, $\Delta$ and the pion as
inputs. The results agree with the experimental values within about
$30\%$ for most of the quantities.

However, the quantization of the spinning collective coordinate
explained above is not a fully satisfactory and consistent one
as we mentioned in the Introduction. Here, we recapitulate the
reasons:
\begin{itemize}
\item
Firstly, the Lagrangian \eqref{L(R,dotR)_RB} is nothing but that
of a spherical rigid body. However, the rotating soliton should
intuitively {\em deform} from the spherical shape, and the
Lagrangian of $R$ should contain terms which automatically
incorporate such deformation.

\item
By identifying the eigenvalues of the rigid body Hamiltonian
\eqref{H_rigidbody} corresponding to the Lagrangian
\eqref{L(R,dotR)_RB}
with the experimental masses of the nucleon and $\Delta$,
we saw that about $8\%$ ($30\%$) of the total mass of the nucleon
($\Delta$) comes from the rotational energy, and that the spinning
velocity at the radius of $1\,\fm$ is nearly (or over) the light
velocity.
Therefore, the validity of the rigid body (i.e., non-relativistic)
approximation for baryons is a non-trivial problem.

\item
The Skyrme field \eqref{U_RB} does {\em not} satisfy the field theory
EOM \eqref{FEOM} even if we use the EOM of $R(t)$ obtained from the
Lagrangian \eqref{L(R,dotR)_RB}:
\begin{equation}
\drv{\bmOmg}{t}=\drv{}{t}\RiRd=0,
\label{REOM}
\end{equation}
which is equivalent to
\begin{equation}
\drv{\bm{\omega}}{t}=\drv{}{t}\dot{R}R^{-1}=0 .
\label{REOM_omega}
\end{equation}
In fact, the EOM \eqref{FEOM} is violated by terms of $O(\bmOmg^2)$:
\begin{align}
&\mbox{LHS of \eqref{FEOM}}\bigr|_{
U(\bmx,t)=\Ucl(\zbmy)}=
\left(\Lcl_a-\frac{1}{\fpiesq}
\left[\Lcl_b,\left[\Lcl_a,\Lcl_b\right]\right]\right)(\zbmy)
\left[\left(\drv{}{t}\RiRd\right)\zbmy\right]_{\!a}
\notag\\
&
+\frac{2}{\fpiesq}\Biggl\{(1-\cos 2F)
\left(F''+\frac{2}{r}F'-\frac{2\sin 2F}{r^2}\right)
+2\sin 2F\,(F')^2
\Biggr\}
\frac{1}{r^3}(\RiRd\zbmy)^2(\zbmy\cdot\bm{\tau})
\notag\\
&
+\left\{1+\frac{4}{\fpiesq}\left[(F')^2
-\frac{1-\cos 2F}{r^2}\right]\right\}
\notag\\
&\qquad\qquad
\times\left\{
\frac{\sin 2F}{2r}\bigl((\RiRd)^2\zbmy\bigr)_a\tau_a
-\frac{1-\cos 2F}{2r^2}
\bigl[\zbmy\times\bigl((\RiRd)^2\zbmy\bigr)\bigr]_a\tau_a
\right\} ,
\label{V_FEOM}
\end{align}
where $\zbmy$ and $\Lcl_a(\zbmy)$ are defined by
\begin{align}
&\zbmy=\zbmy(\bmx,t)=R^{-1}(t)\bmx ,
\label{zbmy}
\\
&\Lcl_a(\zbmy)=-i\Ucl(\zbmy)\Drv{}{\zy_a}\Ucl(\zbmy)^\dagger ,
\label{Lcl}
\end{align}
and we have used the EOM \eqref{FEOM} for $\Ucl$.
(For the derivation of \eqref{V_FEOM}, see Appendix
\ref{app:deriv_FEOM_OSS}, where we present the derivation
\eqref{FEOM_OSS} with relativistic corrections.)
Therefore, the quantization of spinning motion
starting with \eqref{U_RB} is valid only for slowly
rotating baryons with small angular velocity $\bmOmg$.

\end{itemize}
In the subsequent sections, we study relativistic corrections to the
rigid body approximation on the basis of the EOM principle stated in
the Introduction.

\section{Realizing the EOM principle to $\bm{O(\p_t^2)}$}
\label{sec:realizeEOMP}

As we saw in the previous section, the spinning Skyrmion field
given by \eqref{U_RB} does not satisfy our EOM principle for
introducing the collective coordinates: The field theory EOM
\eqref{FEOM} is broken by terms of $O(\bmOmg^2)=O(\p_t^2)$ after using
the EOM of $R(t)$ (see \eqref{V_FEOM}).
Let us therefore try to resolve this $O(\p_t^2)$ breaking of
field theory EOM to shift it to the next higher order of
$O(\p_t^4)$.
Although we are still assuming that the rotation is not so fast, our
analysis in this paper must prove an important step toward the
complete realization of our EOM principle.

\subsection{Improved spinning Skyrmion field}
\label{sec:ISSF}

Our proposal for the improved Skyrmion field with the collective
coordinate $R(t)$ of spinning motion is that all the $R$-dependence is
contained in the argument of $\Ucl$ as in \eqref{U_RB} in the rigid
body approximation:
\begin{equation}
U(\bmx,t)=\Ucl\bigl(\bmy(\bmx,t)\bigr) ,
\label{OSS}
\end{equation}
where $\bmy(\bmx,t)$ is given by
\begin{equation}
\bmy(\bmx,t)=\left[1+A(r)(\dot{R}R^{-1}\bmx)^2+B(r)r^2\Tr(\RiRd)^2
+C(r)r^2(\RiRd)^2\right]\!R^{-1}\bmx
\label{bmy=}
\end{equation}
with $r=\abs{\bmx}$.
Compared with \eqref{U_RB} in the rigid body approximation, we have
introduced terms of $O(\bmOmg^2)$ in the argument of $\Ucl$.
The present $\bmy(\bmx,t)$ is a spinning motion analogue of
\eqref{y(x,t)} for the center-of-mass motion.
The three functions $A(r)$, $B(r)$ and $C(r)$ in \eqref{bmy=} should
be determined to fulfill our EOM principle.
However, it is quite non-obvious at this stage whether the improvement
can be accomplished by \eqref{OSS} and \eqref{bmy=}, namely, whether
we can consistently determine $(A(r),B(r),C(r))$ so that our EOM
principle is satisfied to $O(\bmOmg^2)$.

The form of $\bmy(\bmx,t)$ \eqref{bmy=} is the most general one
containing at most two time-derivatives and satisfying the following
requirements:
\begin{itemize}
\item
$\bmy(\bmx,t)$ is odd under $\bmx\to -\bmx$,
\begin{equation}
\bmy(-\bmx,t)=-\bmy(\bmx,t) ,
\end{equation}
and is even under the time-inversion of $R(t)$:
\begin{equation}
\bmy(\bmx,t)\bigr|_{R(t)\to R(-t)}=\bmy(\bmx,-t) .
\label{y|R(t)->R(-t)}
\end{equation}
\item
$\bmy(\bmx,t)$ has the following property under the constant left and
right $SO(3)$ transformations of $R(t)$:
\begin{equation}
\bmy(\bmx,t)\bigr|_{R(t)\to\calO_\text{real}^{-1}R(t)\calO_\text{iso}}
=\calO_\text{iso}^{-1}\bmy\bigl(\calO_\text{real}\bmx,t\bigr)
\label{bmy_RtoOinvRO}
\end{equation}
This implies that our improved $U(\bmx,t)$ of \eqref{OSS} keeps the
property \eqref{ImpProp}.

\item
$\bmy(\bmx,t)$ does not contain $\ddot{R}(t)$. This is necessary for
the Lagrangian of $R(t)$ to consist of $R$ and $\dot{R}$ without
$\ddot{R}$.
\end{itemize}

As we will see later in Sec.\ \ref{sec:RLagrangian}, the EOM of $R(t)$
corresponding to the improved $U(\bmx,t)$ \eqref{OSS} remains the same
as \eqref{REOM}.
Under the EOM \eqref{REOM}, our Skyrmion field \eqref{OSS} is indeed
spinning both in space and isospace with angular velocities
$\bm{\omega}$ \eqref{omega_i} and $\bmOmg$ \eqref{Omega_a},
respectively. This fact may precisely be stated as follows.
First, $\bmy(\bmx,t)$ of \eqref{bmy=} has the following property:
\begin{equation}
\left.
\left(\Drv{}{t}+(\dot{R}R^{-1}\bmx)_i\Drv{}{x_i}
\right)\bmy(\bmx,t)\right|_{R\text{-EOM}}=0 ,
\label{EDbmy=0}
\end{equation}
where ``$|_{R\text{-EOM}}$" means
``upon using the EOM \eqref{REOM} of $R(t)$''.
Since $U(\bmx,t)$ \eqref{OSS} is a function of $\bmy$ only, it obeys
the same equation as \eqref{EDbmy=0}:
\begin{equation}
\left.
\left(\Drv{}{t}+(\bm{\omega}\times\bmx)_i\Drv{}{x_i}
\right)U(\bmx,t)\right|_{R\text{-EOM}}=0 .
\end{equation}
This implies that $U(\bmx,t)$ is spinning in real space with angular
velocity $\bm{\omega}$ when the EOM \eqref{REOM} holds.
Next, $\bmy(\bmx,t)$ also obeys
\begin{equation}
\left.\Drv{\bmy(\bmx,t)}{t}\right|_{R\text{-EOM}}
=-\RiRd\bmy=\bmOmg\times\bmy .
\label{delydelt}
\end{equation}
This together with the hedgehog property \eqref{hedgehogP} means that
$U(\bmx,t)$ is spinning in isospace with angular velocity
$\bmOmg$.

\begin{figure}[htbp]
\centering
\includegraphics[scale=0.5]{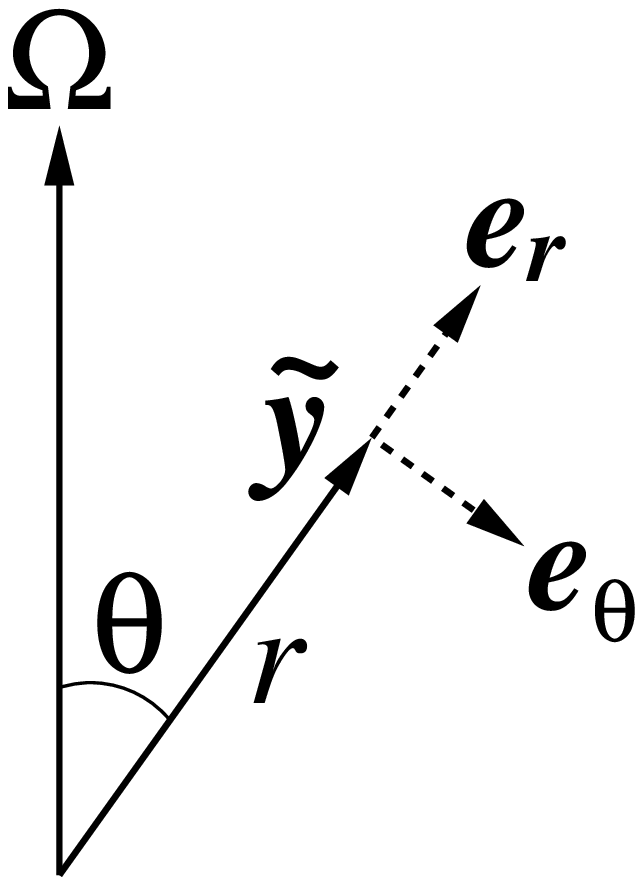}
\caption{
Polar coordinate $(r,\theta)$ for $\zbmy$, and the unit vectors
$(\bm{e}_r,\bm{e}_\theta)$ in \eqref{uinp}.}
\label{polar}
\end{figure}

Finally in this subsection, we consider the geometrical meaning of
$\bmy(\bmx,t)$ \eqref{bmy=}, and, in particular, the spatial
shape represented by our spinning Skyrmion \eqref{OSS}.
Below we assume that $r^2\bmOmg^2$ is sufficiently small.
Let $\bm{u}$ be the displacement vector, namely, the
difference of $\bmy$ \eqref{bmy=} and $\zbmy=R^{-1}\bmx$ \eqref{zbmy}:
\begin{equation}
\bmy=\zbmy + \bm{u} .
\end{equation}
Taking, in the $\zbmy$-space, the polar coordinate system
$(r\!=\!\abs{\zbmy}\!=\!\abs{\bmx},\theta,\varphi)$ with $z$-axis in
the $\bmOmg$ direction, $\bm{u}$ has the following decomposition
in terms of the unit vectors $\bm{e}_r=\zbmy/r$ and $\bm{e}_\theta$
(see Fig.\ \ref{polar}):
\begin{equation}
\bm{u}=-\left\{\frac23\Bigl[Y(r)+\bigl(A(r)-C(r)\bigr)P_2(\cos\theta)
\Bigr]\bm{e}_r+C(r)\sin\theta\cos\theta\,\bm{e}_\theta\right\}
r^3\bmOmg^2 ,
\label{uinp}
\end{equation}
where the function $Y(r)$ is defined by
\begin{equation}
Y(r)=-A(r)+3 B(r) + C(r) ,
\label{defY}
\end{equation}
and $P_2$ is a Legendre polynomial:
\begin{equation}
P_2(z)=\frac12\left(3z^2-1\right) .
\end{equation}
Note that the functions $(A,B,C)$ appear in $\bm{u}$ as another three
independent combinations; $Y$, $A-C$ and $C$, which also appear in
other places below.
The decomposition \eqref{uinp} of $\bm{u}$ is restated as the
following relation between the polar coordinate
$(\abs{\bmy},\theta_{\bmy},\varphi)$ of $\bmy$ and
$(r,\theta,\varphi)$ of $\zbmy$ (the $\varphi$ coordinate is common
since there is no $\bm{e}_\varphi$ term in \eqref{uinp}):
\begin{align}
\abs{\bmy}&=\left\{
1-\frac23\Bigl[Y(r)+\bigl(A(r)-C(r)\bigr)P_2(\cos\theta)
\Bigr]r^2\bmOmg^2\right\}r ,
\label{absbmy=}
\\
\tan\theta_{\bmy}&=\left(1-C(r)r^2\bmOmg^2\right)\tan\theta ,
\label{tan=(...)tan}
\end{align}
where $\theta_{\bmy}$ is the angle between $\bmy$ and $\bmOmg$.

The (classical) shape of our spinning Skyrmion \eqref{OSS} in
$\bmx$-space for a given value of $\bmOmg^2$ may have various
definitions; for example, surfaces of constant energy density,
constant baryon number density, and so on.\footnote{
The isoscalar quadrupole moment operator, which vanishes identically
in the rigid body approximation, can be non-trivial due to present
relativistic correction and is given by \eqref{ISQM} in Appendix
\ref{app:AE}. However, its nucleon expectation value is equal to zero.
}
Here, noting that
$\Ucl(\bmy)$ represents a spherically symmetric object in
$\bmy$-space, let us take the simplest one; surfaces of
constant $\abs{\bmy}$. From \eqref{absbmy=}, the surface of
$\abs{\bmy}=a\,(=\text{const.})$ is mapped to the $\bmx$-space as
\begin{equation}
\frac{a}{r}=1+\Bigl[\bigl(A(a)-C(a)\bigr)\sin^2\theta -2B(a)
\Bigr]a^2\bmOmg^2
+O(a^4\bmOmg^4) .
\label{riny}
\end{equation}
Here, we have $r=\abs{\bmx}$, and $\pi-\theta$ is the angle between
$\bm{\omega}$ and $\bmx$. Eq.\ \eqref{riny} represents an ellipse on
the $(r,\theta)$ plane, and hence our spinning Skyrmion \eqref{OSS}
has a shape of spheroid, as is intuitively expected.

\subsection{Differential equations and boundary conditions of
  $\bm{(A,B,C)}$}
\label{sec:DEBC}

The three functions $(A,B,C)$ in \eqref{bmy=} must be determined in
such a way that the field theory EOM \eqref{FEOM} with the spinning
Skyrmion field \eqref{OSS} substituted holds to $O(\bmOmg^2)$
upon using the EOM of the collective coordinate $R(t)$.
Namely, the contribution of the $\bmOmg^2$ terms in \eqref{bmy=} to
the field theory EOM must cancel the $\bmOmg^2$ terms in
\eqref{V_FEOM}.

After a tedious but straightforward calculation by using
\begin{equation}
L_0(\bmx,t)=\Drv{y_a(\bmx,t)}{t}\Lcl_a(\bmy),
\qquad
L_i(\bmx,t)=\Drv{y_a(\bmx,t)}{x_i}\Lcl_a(\bmy) ,
\label{LmubyLcl}
\end{equation}
with $\Lcl_a(\bmy)=-i\Ucl(\bmy)(\p/\p y_a)\Ucl(\bmy)^\dagger$ defined
by \eqref{Lcl}, we get, instead of \eqref{V_FEOM}, the following
result (we summarize the derivation in Appendix
\ref{app:deriv_FEOM_OSS}):
\begin{align}
&\mbox{LHS of \eqref{FEOM}}\bigr|_{U(\bmx,t)=\Ucl(\bmy)}
=\left(\Lcl_a-\frac{1}{\fpiesq}
\bigl[\Lcl_b,[\Lcl_a,\Lcl_b]\bigr]\right)(\bmy)
\left[\left(\drv{}{t}\RiRd\right)\bmy\right]_a
\notag \\
&\quad
+r^2\Tr(\RiRd)^2(\bmy\cdot\bm{\tau})\times\EQ_1
+(\RiRd\bmy)^2(\bmy\cdot\bm{\tau})\times \EQ_2
\notag \\
&\quad
+r^2\bigl[(\RiRd)^2\bmy\bigr]\cdot\bm{\tau}\times \EQ_3
+r\bigl[\bmy\times(\RiRd)^2\bmy\bigr]
\cdot\bm{\tau}\times \EQ_4
+O(\p_t^4),
\label{FEOM_OSS}
\end{align}
where $\EQ_n$ ($n=1,2,3,4$) are linear in $(A,B,C)$ and their first
and second derivatives with respect to $r$ with coefficients given in
terms of $F$ and its $r$-derivatives. The concrete expressions of
$\EQ_n$ are very lengthy, and they are presented in Appendix
\ref{app:EQ_1234}.
As we have already mentioned in Sec.\ \ref{sec:ISSF}, and as we will
see in the next section, the EOM of the collective coordinate $R(t)$
derived from its Lagrangian, which is obtained by substituting
\eqref{OSS} into the Skyrme Lagrangian density \eqref{calL_Skyrme} and
carrying out the space integration, remains unchanged from
\eqref{REOM} in the rigid body approximation.
This implies that the first term on the RHS of \eqref{FEOM_OSS}
vanishes upon using the EOM of $R(t)$, and our EOM principle is
satisfied to $O(\bmOmg^2)$ if the three functions $(A,B,C)$
satisfy four differential equations,
\begin{equation}
\EQ_n=0,\quad(n=1,2,3,4) .
\label{EQ_n=0}
\end{equation}
This is apparently overdetermined, but, fortunately, $\EQ_3$ and
$\EQ_4$ are not independent. We have
\begin{equation}
\EQ_3=2\cos F\times\EQ_{34},\quad
\EQ_4=-2\sin F\times\EQ_{34},
\label{EQ3EQ4}
\end{equation}
with $\EQ_{34}$ given by \eqref{EQ34}.
Therefore, $(A,B,C)$ are determined by solving three inhomogeneous
differential equations of second order, $\EQ_1=\EQ_2=\EQ_{34}=0$.

As we can guess from the fact that $Y(r)$ \eqref{defY} is the
coefficient of the lowest mode $1=P_0(\cos\theta)$ in \eqref{absbmy=},
we can extract the differential equation for a single function $Y(r)$
by taking a suitable linear combination of $\EQ_{1,2,3}$.
In fact, the special combination,
\begin{equation}
\EQ_Y=-3\,\EQ_1 + \EQ_2 - \EQ_3 .
\label{defEQY}
\end{equation}
consists only of $Y(r)$ and its derivatives as given in \eqref{EQY}.
Note also that $B(r)$ and its derivatives are missing from $\EQ_2$
\eqref{EQ2} and $\EQ_{34}$ \eqref{EQ34}.
Therefore, for practical purposes, it is convenient to solve $\EQ_Y=0$
for $Y(r)$, and $\EQ_2=\EQ_{34}=0$ for $(A(r),C(r))$.

For solving the differential equations for $(A,B,C)$, we have to
specify the boundary conditions of the three functions at $r=0$ and
$r=\infty$. First, substituting the approximate expression of
$F(r)$ near the origin given by \eqref{F_origin} into $\EQ_2$,
$\EQ_{34}$ and $\EQ_Y$, we obtain the following differential equations
for $(A,C)$ and $Y$ valid for $r\sim 0$:
\begin{align}
&(1+8\mu^2)r^2\drv{^2A}{r^2}-4\mu^2r^2\drv{^2C}{r^2}
+8(1+7\mu^2)r\drv{A}{r}-12\mu^2r\drv{C}{r}=0 ,
\notag\\
&(1+4\mu^2)r^2\drv{^2C}{r^2}-8\mu^2r\drv{A}{r}
+8(1+5\mu^2)r\drv{C}{r}
-4(1+14\mu^2)A +(10+56\mu^2)C=1-4\mu^2,
\notag\\
&r^2\drv{^2 Y}{r^2}+8r\drv{Y}{r}+10Y
=\frac{1-4\mu^2}{1+8\mu^2} ,
\label{DE_ABC_origin}
\end{align}
with dimensionless $\mu$ defined by $\mu=\kappa/e\fpi$.
{}From this we find that $(A,C)$ and $Y$ near the origin is generically
given in terms of a particular solution and six independent modes:
\begin{align}
\Pmatrix{A(r)\\ C(r)}
&=\frac{1-4\mu^2}{10(1+8\mu^2)}\Pmatrix{4\mu^2\\ 1+4\mu^2}
\!+\!c_0\Pmatrix{5+28\mu^2\\ 2+28\mu^2}
\!+\!c_2\Pmatrix{0\\1}\frac{1}{r^2}
\!+\!c_5\Pmatrix{-6\mu^2\\ 1+4\mu^2}\frac{1}{r^5}
\!+\!c_7\Pmatrix{5\\ 2}\frac{1}{r^7},
\notag\\
Y(r)&=\frac{1-4\mu^2}{10(1+8\mu^2)}+c_2^Y\frac{1}{r^2}
+c_5^Y\frac{1}{r^5},
\qquad (r\sim0),
\label{ABCorigin}
\end{align}
where $c_n$ ($n=0,2,5,7$) and $c_n^Y$ ($n=2,5$) are constants.
Each term on the RHSs of \eqref{ABCorigin}, for example,
$\smallPmatrix{5\\ 2}(1/r^7)$, means a series in
$r^2$ starting with this term.

On the other hand, plugging the $r\to\infty$ behavior of $F(r)$
given by \eqref{F_infty} into $\EQ_2$, $\EQ_{34}$ and $\EQ_Y$,
we obtain the differential equations for $(A,C)$ and $Y$
approximated near the infinity.
They are lengthy and are given by \eqref{DE_ABC_infty} in
Appendix \ref{app:EQ_1234} by using the dimensionless variable
$s=\mpi r$.
The Skyrme term does not contribute to these differential equations.
{}From \eqref{DE_ABC_infty}, we find that $(A,C)$ and $Y$ near the
infinity are given as the following sum of a particular solution and
six independent modes:
\begin{align}
\Pmatrix{A(r)\\ C(r)}
&=-\frac{1}{2s}\Pmatrix{1\\ 1}
+d_2\Pmatrix{1\\ 1}\frac{1}{s^2}+d_3\Pmatrix{1\\ 2}\frac{1}{s^3}
+\left\{
f_2\Pmatrix{1\\ 1}\frac{1}{s^2}+f_3\Pmatrix{3\\ 2}\frac{1}{s^3}
\right\}e^{2s} ,
\notag\\
Y(r)&=\frac{1}{2s^2}+d_3^Y\frac{1}{s^3}+f_3^Y\frac{e^{2s}}{s^3},
\qquad(r\to\infty),
\label{ABCinfty}
\end{align}
where $d_{2,3}$, $f_{2,3}$, $d_3^Y$ and $f_3^Y$ are constants,
and each term on the RHSs shows the first term of the series in
$1/s$. Concrete expressions keeping terms up to $1/s^5$ for the power
parts are given in \eqref{DE_ABC_infty_series} in Appendix
\ref{app:EQ_1234}.

For solving the differential equations for $(A,B,C)$ globally in the
range $0\le r<\infty$, we impose the following boundary conditions in
terms of the coefficients appearing in \eqref{ABCorigin} and
\eqref{ABCinfty}:
\begin{align}
&c_5=c_5^Y=c_7=0,
\label{BC_origin}
\\
&f_2=f_3=f_3^Y=0.
\label{BC_infty}
\end{align}
Namely, we choose the ``mildest'' boundary conditions both at the
origin and the infinity.
As we will see later, the most important functional of $(A,B,C)$
is $\calJ$ given by \eqref{calJ}. It is the coefficient of the
$\bmOmg^4$ term of the Lagrangian \eqref{L_R} of $R(t)$ and
hence appears in various conserved charge operators such as
Hamiltonian, angular momentum and isospin.
Owing to our boundary conditions, \eqref{BC_origin} and
\eqref{BC_infty}, the $r$-integration for $\calJ$ is convergent at
both ends, $r=0$ and $r=\infty$.
This is the case also for other physical quantities which we will
discuss later in Sec.\ \ref{sec:SPN}.
Moreover, the deformation vector $\bm{u}(\bmx,t)$
\eqref{uinp} is non-divergent at $r=0$. It is also finite in the limit
$r\to\infty$ if we fix $r^2\bmOmg^2=\text{finite}$.
These properties look natural for our improvement to make sense.

Having fixed the boundary conditions for $(A,B,C)$, we can numerically
solve the differential equations for them by the shooting method.
Namely, setting the boundary condition \eqref{BC_origin} at $r=0$,
we tune the three parameters $(c_0,c_2,c_2^Y)$ in
\eqref{ABCorigin} so that the condition \eqref{BC_infty} at $r=\infty$
is satisfied. This task of solving the differential equations for
$(A,B,C)$ will be done in Sec.\ \ref{sec:DetfpieABC} after we
obtain the Lagrangian and the Hamiltonian for $R(t)$.

Finally, we add that our EOM principle implies that the spinning
Skyrmion field \eqref{OSS} with a constant $\bmOmg$ is nothing but a
classical spinning solution of the Skyrme field theory up to
$O(\bmOmg^4)$.

\section{Lagrangian of $\bm{R(t)}$}
\label{sec:RLagrangian}

The Lagrangian of $R(t)$ in our formalism with relativistic correction
is given, as in the rigid body approximation of \eqref{L(R,dotR)_RB},
by substituting the improved Skyrmion field \eqref{OSS} into the
Lagrangian density \eqref{calL_Skyrme} and carrying out the space
integration.
The corrections we have added in \eqref{bmy=} contain two
time-derivatives, and hence we can consider terms with four
time-derivatives in the Lagrangian.
We find that it is given by
\begin{equation}
L(R,\dot{R})=\int\! d^3x\,\calL\bigr|_{U(\bmx,t)=\Ucl(\bmy)}
=-\Mcl +\frac12\calI\bmOmg^2
+\frac14\calJ\bmOmg^4 ,
\label{L_R}
\end{equation}
where the rest mass $\Mcl$ and the moment-of-inertia $\calI$ are the
same as in the rigid body approximation, \eqref{L(R,dotR)_RB}, and the
coefficient $\calJ$ of the newly added $\bmOmg^4$ term is given by
\begin{align}
\calJ&=\frac{4\pi\fpi^2}{15}\int_0^\infty\!dr\,r^4\sin^2 F
\notag\\
&\qquad\times\left\{rZ'+5Z-C
+\frac{4}{\fpiesq}\left[
\frac{\sin^2 F}{r^2}\left(rZ'+3Z+2C\right)
-(F')^2\left(rZ'+Z+C\right)\right]\right\} ,
\label{calJ}
\end{align}
with $Z(r)$ defined by\footnote{
The function $Z(r)$ is extracted from the coefficient of $\bm{e}_r$ in
\eqref{uinp} by carrying out the $\cos\theta$-integration with weight
function $\sin^2\theta$:
$$
\frac12\int_{-1}^1\!d(\cos\theta)\Bigl[
Y+(A-C)P_2(\cos\theta)\Bigr]\sin^2\theta=\frac25Z .
$$
}
\begin{equation}
Z=-2A+5B+2C=\frac53 Y-\frac13(A-C) .
\label{Z}
\end{equation}
The Lagrangian \eqref{L_R} should be regarded as the first three terms
of the rotational motion counterpart of the relativistic Lagrangian of
the center-of-mass coordinate $\bm{X}(t)$;
$-\Mcl\sqrt{1-\bm{V}^2}=-\Mcl+(1/2)\Mcl\bm{V}^2+(1/8)\Mcl\bm{V}^4
+\ldots$.
The EOM of $R(t)$ derived from the Lagrangian \eqref{L_R} reads
\begin{equation}
\drv{}{t}\left[\left(\calI +\calJ\bmOmg^2\right)\bmOmg
\right]=0 .
\end{equation}
This generically implies $\dot{\bmOmg}=0$. Namely, the EOM of
$R(t)$ remains unchanged from \eqref{REOM} in the rigid body
approximation, which we already used deriving the differential
equations for $(A,B,C)$ in Sec.\ \ref{sec:DEBC}.

Our result \eqref{L_R}, in particular, $\calJ$ of \eqref{calJ}, can be
obtained by a rather lengthy calculation with use of \eqref{LmubyLcl}
for $L_\mu$. The outline of the derivation of \eqref{calJ} is given in
Appendix \ref{app:calc_calJ}. Here, we explain a number of points
used in deriving \eqref{L_R} and \eqref{calJ}.

First, note that Lagrangian \eqref{L_R} consists only of
the angular velocity $\bmOmg$; $\ddot{R}(t)$ and hence $\dot{\bmOmg}$
are missing.
It is due to this fact that the EOM of $R(t)$ remains essentially the
same as in the rigid body approximation. The reason why the higher
time-derivative of $R(t)$ does not appear in the Lagrangian is as
follows.
Since the Lagrangian of $R(t)$ has an invariance under
\eqref{RtoOinvRO} due to the property \eqref{bmy_RtoOinvRO} and hence
\eqref{ImpProp}, and since $\bmy(\bmx,t)$ \eqref{bmy=} does not
contain $\ddot{R}$ nor terms with a single $\RiRd$, the possible terms
in the Lagrangian of $R(t)$ containing $\ddot{R}$ and at most four
time-derivatives in total are $\Tr\bigl(\RiRd\,(d/dt)\RiRd\bigr)$
and $\Tr\bigl((\RiRd)^2(d/dt)\RiRd\bigr)$.
The origin of these terms is the part quadratic in $L_0$
in the Skyrme model Lagrangian density \eqref{calL_Skyrme}.
However, neither of these two terms can actually exist:
The former term with three time-derivatives cannot appear in
the Lagrangian owing to another property \eqref{y|R(t)->R(-t)} of
$\bmy(\bmx,t)$ (this term is in any case a time-derivative term
and can be discarded even if it does exist).
The latter term vanishes identically since $\RiRd$ is an
anti-symmetric matrix.

Now, the Lagrangian of $R(t)$ takes symbolically the form
$L=1+\bmOmg^2 +\bmOmg^4$.
To consider the coefficients of the three terms, we divide the
Lagrangian into the ``kinetic part'' $T$ (the part with $L_0$) and
the ``potential part'' $V$ (the part without $L_0$), and write
$L=T-V$ with
\begin{align}
T&=\int\!d^3x\,\tr\left.\left\{
\frac{\fpi^2}{16}L_0^2-\frac{1}{16e^2}[L_0,L_i]^2\right\}
\right|_{U(\bmx,t)=\Ucl(\bmy)} ,
\label{KP}
\\
V&=\int\!d^3x\,\tr\left.\left\{
\frac{\fpi^2}{16}L_i^2-\frac{1}{32e^2}[L_i,L_j]^2
-\frac{\fpi^2}{8}\mpi^2(U-\bm 1_2)\right\}
\right|_{U(\bmx,t)=\Ucl(\bmy)} .
\label{PP}
\end{align}
Then, $T$ and $V$ have the following expressions:
\begin{align}
T&=\frac12\calI\bmOmg^2+\frac14\calJ_1\bmOmg^4+
O(\p_t^6),
\label{Tpart}\\
V&=\Mcl-\frac12\Delta\calI\bmOmg^2-\frac14\calJ_2\bmOmg^4
+O(\p_t^6),
\label{Vpart}
\end{align}
and hence
\begin{equation}
L=-\Mcl+\frac12\left(\calI+\Delta\calI\right)\bmOmg^2
+\frac14\left(\calJ_1+\calJ_2\right)\bmOmg^4 .
\label{L=1+Omg^2+Omg^4}
\end{equation}
The origin of each term in \eqref{Tpart} and \eqref{Vpart} is
explained as follows. Since $\bmy=R^{-1}\bmx+\bm{u}$ of \eqref{bmy=}
is the sum of the rigid body term $R^{-1}\bmx$ and the correction term
$\bm{u}$ which are quadratic in $\RiRd$, the coefficients of the
lowest order terms in \eqref{Tpart} and \eqref{Vpart}, namely, $\calI$
and $\Mcl$, are the same as those in the rigid body approximation.
Next, the $\calJ_1\bmOmg^4$ and the $\Delta\calI\bmOmg^2$
terms are the leading terms due to our relativistic correction
$\bm{u}$, and hence they are linear in $(A,B,C)$. Finally, the
$\calJ_2\bmOmg^4$ term in \eqref{Vpart} is quadratic in $\bm{u}$
and therefore is quadratic in $(A,B,C)$.
One might wonder that, if we have added the $O(\p_t^4)$ terms to
$\bmy$, they would also contribute to the $\bmOmg^4$ term
in $V$. However, this is not the case as we will see below.

By a simple scaling argument, we can get useful relationships among
the coefficients in \eqref{L=1+Omg^2+Omg^4}. Let us replace $(A,B,C)$
in $\bmy$ of \eqref{bmy=} by $\lambda(A,B,C)$ with $\lambda$ being a
constant parameter, and denote this modified $\bmy$ by $\bmy_\lambda$;
namely, $\bmy_\lambda=R^{-1}\bmx+\lambda\bm{u}$.
Then, using the fact that $\Delta\calI$ and $\calJ_1$ are linear in
$(A,B,C)$, while $\calJ_2$ is quadratic in them,
the Lagrangian of $R(t)$ corresponding to
$U(\bmx,t)=\Ucl\bigl(\bmy_\lambda(\bmx,t)\bigr)$ reads, instead of
\eqref{L=1+Omg^2+Omg^4}, as follows:
\begin{equation}
L_\lambda=-\Mcl
+\frac12\left(\calI+\lambda\Delta\calI\right)\bmOmg^2
+\frac14\left(\lambda\calJ_1+\lambda^2\calJ_2\right)\bmOmg^4 .
\label{L_lambda}
\end{equation}
At this point we should recall our EOM principle of introducing
collective coordinates; $U(\bmx,t)=\Ucl(\bmy_{\lambda=1})$ satisfies
the field theory EOM, namely, it is an extrema of the field theory
action if $R(t)$ is subject to its EOM, $\dot{\bmOmg}=0$.
This implies, in particular, that $L_\lambda$ \eqref{L_lambda} is
stationary at $\lambda=1$ for any constant $\bmOmg$:
\begin{equation}
0=\left.\drv{}{\lambda}L_\lambda\right|_{\lambda=1}
=\frac12\Delta\calI\,\bmOmg^2
+\frac14\left(\calJ_1+2\calJ_2\right)\bmOmg^4 ,
\qquad\mbox{(for ${}^\forall\bmOmg$)} .
\end{equation}
Therefore, we get
\begin{equation}
\Delta\calI=0,
\qquad
\calJ_1+2\calJ_2=0.
\label{Virial}
\end{equation}
By the same rescaling argument (by using another parameter), we can
show that the addition of $O(\p_t^4)$ terms to $\bm{u}$ does not
affect the coefficient of $\bmOmg^4$ in \eqref{L=1+Omg^2+Omg^4}
as we mentioned above.\footnote{
The same kinds of relationships also hold in the case of the
relativistic correction to the center-of-mass motion described by
\eqref{y(x,t)}, \eqref{L_X} and \eqref{EOM_X}: In the Lagrangian
\eqref{L_X}, $c_2$ does not affect the coefficient of the $\dot{X}^2$
term, the relation $c_2+2(-c_2^2)=0$ holds for the coefficient
$c_2-c_2^2$ of $\dot{X}^4$ (since we have $c_2=1/2$), and $c_4$ does
not enter the coefficient of the $\dot{X}^4$ term.
}

Owing to the second relation of \eqref{Virial}, the coefficient
$\calJ$ of $\bmOmg^4$ in \eqref{L=1+Omg^2+Omg^4} can be expressed
only in terms of $\calJ_1$:
\begin{equation}
\calJ=\calJ_1+\calJ_2=\frac12\,\calJ_1
\label{calJ=(1/2)calJ_1}
\end{equation}
This is very beneficial for obtaining $\calJ$ since $\calJ_2$, which
is quadratic in $(A,B,C)$, is harder to evaluate.
In Appendix \ref{app:calc_calJ}, we present the derivation of
$\calJ$ \eqref{calJ} by using \eqref{calJ=(1/2)calJ_1}.

Some comments are in order concerning our results \eqref{L_R} and
\eqref{calJ} and the boundary conditions \eqref{BC_origin} and
\eqref{BC_infty} we have chosen for $(A,B,C)$.
First, from the boundary behaviors of $F(r)$ given by \eqref{F_origin}
and \eqref{F_infty}, the $r$-integration of $\calJ$ \eqref{calJ}
is approximated near $r=0$ and $r=\infty$ by
$\int_{r\sim 0}\!dr\,r^6\left[rZ'+5Z-C+(2\kappa/e\fpi)^2(2Z+C)
\right]$ and $\int^{r\sim\infty}\!dr\,r^2 e^{-2\mpi r}(rZ'+5Z-C)$,
respectively. They are both convergent due to the boundary conditions
$c_7=0$ at $r=0$ and \eqref{BC_infty} at $r=\infty$.

Our second comment is on the necessity of the non-zero pion mass.
If the pion mass $\mpi$ is zero, the exponential falloff
\eqref{F_infty} of $F(r)$ near $r=\infty$ is changed to the power one
$F(r)\sim 1/r^2$. Correspondingly, choosing the mildest boundary
condition at $r=\infty$, we have the following leading behaviors near
$r=\infty$; $(A,C)\sim (-3/4,-1/2)$ and $Y\sim 1/4$.
This implies that the $r$-integration \eqref{calJ} for $\calJ$ is
linearly divergent at $r=\infty$ in the case of $\mpi=0$.
This is the case also for relativistic corrections to
other physical quantities which we discuss in Sec.\ \ref{sec:SPN}.
Therefore, the inclusion of non-zero pion mass is inevitable for our
analysis in contrast to the case of the rigid body approximation
\cite{ANW}.
(On the other hand, the qualitative behaviors of the various
quantities near the origin $r=0$ are insensitive to $\mpi$.)

Next, as described in Appendix
\ref{app:calc_calJ}, the expression \eqref{calJ} for $\calJ$ has been
obtained by changing the variables of integration in \eqref{L_R} from
$\bmx$ to $\bmy$, and carrying out the integration over the
solid-angle of $\bmy$. Therefore, $r$ in \eqref{calJ} is
the length of $\bmy$; $r=\abs{\bmy}$.
This way of switching to the $\bmy$-integration is easier than to
consider the original $\bmx$-integration.
If we persist in carrying out the $\bmx$-integration, we
obtain the expression of $\calJ$ which differs from \eqref{calJ} by a
surface term:
\begin{equation}
\calJ\bigr|_{\text{by $\bmx$-integration}}=\mbox{eq.\,\eqref{calJ}}
-\frac{4\pi\fpi^2}{15}\left[r^5\sin^2 F\left\{
1+\frac{4}{\fpiesq}\left((F')^2+\frac{\sin^2 F}{r^2}\right)
\right\}Z\right]^{r=\infty}_{r=0} ,
\label{eq:calJinx}
\end{equation}
with $r=\abs{\bmx}$ on the RHS. From \eqref{F_origin} and
\eqref{F_infty} for $F(r)$, the surface terms of \eqref{eq:calJinx}
at $r=0$ and $r=\infty$ are (up to constants)
$\lim_{r\to 0}r^7 Z(r)$ and $\lim_{r\to\infty}r^3e^{-2\mpi r}Z(r)$ ,
respectively. Fortunately, they both vanish safely owing to the
boundary conditions \eqref{BC_origin} and \eqref{BC_infty} of
$(A,B,C)$.

Our last comment is on $\Delta\calI$ in \eqref{L=1+Omg^2+Omg^4}.
Explicit calculation by using the $\bmy$-integration and the EOM
\eqref{DEF} of $F(r)$ leads to the following expression of
$\Delta\calI$ as a surface term:
\begin{align}
\Delta\calI&=\frac{2\pi\fpi^2}{3}\left[r^5\left\{
(F')^2-\frac{2\sin^2 F}{r^2}+\frac{4}{\fpiesq}\frac{\sin^2F}{r^2}
\left(2(F')^2-\frac{\sin^2 F}{r^2}\right)
\right\}Y\right]^{r=\infty}_{r=0}
\notag\\
&=\frac{2\pi\fpi^2}{3}\left\{
\kappa^2\left(1-\frac{4\kappa^2}{\fpiesq}\right)
\lim_{r\to 0}r^5 Y(r)
+a^2\mpi^2\lim_{r\to\infty}r^3 e^{-2\mpi r}Y(r)\right\} .
\end{align}
This also vanishes owing to the boundary conditions $c_5^Y=f_3^Y=0$
of $Y(r)$.

\section{$\bm{\fpi}$ and the static properties of nucleons}
\label{sec:detfpi}

In this section, starting with the Lagrangian \eqref{L_R} of $R(t)$
with the relativistic correction term $(1/4)\calJ\bmOmg^4$,
we first determine the parameters $(\fpi,e)$ of the Skyrme model from
the masses of the lightest baryons, the nucleon and $\Delta$, as well
as $\mpi$. This task includes the determination of the functions
$F(r)$ and the newly introduced $(A(r),B(r),C(r))$.
Then, we calculate various static properties of nucleons and compare
them with the experimental values and also with the results of
\cite{AN} without relativistic correction.

\subsection{Hamiltonian and spin/isospin operators}

Let us consider the standard canonical quantization of the system
described by the Lagrangian \eqref{L_R} with dynamical variable
$R(t)\in SO(3)$. For this, we take three independent variables
$\xi^A(t)$ ($A=1,2,3$) (for example, the Euler angles) which
parametrize the $SO(3)$ matrix $R(t)$; $R=R\bigl(\xi^A(t)\bigr)$.
Then, we have
\begin{equation}
\Omega_a=\frac12\epsilon_{abc}(\RiRd)_{bc}
=\dot{\xi}^A\varpi_{Aa} ,
\qquad
\bmOmg^2=g_{AB}(\xi)\dot{\xi}^A\dot{\xi}^B ,
\end{equation}
with $\varpi_{Aa}$ and $g_{AB}(\xi)$ defined by
\begin{equation}
\left(R(\xi)^{-1}\Drv{}{\xi^A}R(\xi)\right)_{ab}
=\varpi_{Ac}(\xi)\,\epsilon_{abc} ,
\qquad
g_{AB}(\xi)=\varpi_{Aa}(\xi)\varpi_{Ba}(\xi) .
\end{equation}
The canonical momentum $\pi_A$ conjugate to $\xi^A$ is
\begin{equation}
\pi_A=\Drv{L}{\dot{\xi}^A}
=\left(\calI+\calJ\bmOmg^2\right)g_{AB}(\xi)\dot{\xi}^B ,
\end{equation}
and the Hamiltonian of the present system is given by
\begin{equation}
H=\pi_A\dot{\xi}^A-L
=\Mcl+\frac12\calI\bmOmg^2+\frac34\calJ\bmOmg^4 .
\label{H}
\end{equation}

For identifying the eigenvalues of the Hamiltonian with the masses of
the baryons, we have to relate $\bmOmg^2$ with the isospin $\bm{I}$
and spin $\bm{J}$, which are the Noether charges corresponding to the
symmetry transformations \eqref{RtoOinvRO}.
The isospin operator is given by
\begin{equation}
I_a=\left(\calI+\calJ\bmOmg^2\right)\Omega_a ,
\label{I_a=}
\end{equation}
and the canonical commutation relation
$\left[\xi^A(t),\pi_B(t)\right]=i\delta^A_B$ leads to
\begin{equation}
\left[I_a,I_b\right]=i\epsilon_{abc}I_c ,
\qquad
\left[I_a,R_{ib}\right]=i\epsilon_{abc}R_{ic} .
\end{equation}
Similarly, the spin operator is
\begin{equation}
J_i=\left(\calI+\calJ\bmOmg^2\right)\omega_i=-R_{ia}I_a ,
\end{equation}
and satisfies
\begin{equation}
\left[J_i,J_j\right]=i\epsilon_{ijk}I_k ,
\qquad
\left[J_i,R_{ja}\right]=i\epsilon_{ijk}R_{ka} ,
\qquad
\left[I_a,J_i\right]=0 .
\end{equation}
Then, the equation relating $\bmOmg^2$ with the spin and the isospin
is
\begin{equation}
\bm{I}^2=\bm{J}^2=\left(\calI+\calJ\bmOmg^2\right)^2\bmOmg^2 .
\label{I^2=J^2=}
\end{equation}

The Hamiltonian and the (iso)spin charges above have been obtained
from the Lagrangian \eqref{L_R} of $R(t)$. One may, however, wonder
whether these conserved charges in the quantum mechanical system of $R$
agree with those from the Skyrme field theory, namely, whether $H$,
$I_a$ and $J_i$ above agree with those obtained by substituting
$U(\bmx,t)=\Ucl(\bmy)$ \eqref{OSS} into the corresponding Noether
charges from the Lagrangian density \eqref{calL_Skyrme} of the
Skyrme model. This agreement indeed holds owing to our EOM principle
(see Appendix \ref{app:equiv_NoetherQ} for a proof).
Furthermore, by using the equality of the isospins $I_a$ obtained
in two ways, we can get the expression \eqref{calJ} of
$\calJ$ directly from the calculation in the field theory side without
relying on the relation \eqref{calJ=(1/2)calJ_1}.
This is explained in Appendix \ref{app:AE}
(see the paragraph below \eqref{(I+JOmg^2)^-1=}).

\subsection{Determining $\bm{(\fpi,e)}$ and $\bm{(A,B,C)}$
from $\bm{(M_N,M_\Delta,\mpi)}$}
\label{sec:DetfpieABC}

Now, let us consider determining $(\fpi,e)$ from the Hamiltonian
\eqref{H} together with the relation \eqref{I^2=J^2=} by
taking the masses of the nucleon, $\Delta$ and the pion,
\begin{equation}
M_N=939\,\MeV ,\quad M_\Delta=1232\,\MeV ,\quad
\mpi=138\,\MeV ,
\end{equation}
as inputs.
Concrete procedure is as follows. We rewrite the differential
equations for $F$ and $(A,B,C)$ and the integrations for $\calI$
\eqref{calI} and $\calJ$ \eqref{calJ} in terms of the dimensionless
variable $\rdl=e\fpi r$. Then, the differential equations and the
integrations for the dimensionless quantities
$(\Mcldl,\calIdl,\calJdl)$ defined by
\begin{equation}
\Mcldl=\frac{e}{\fpi}\Mcl ,\quad
\calIdl=e^3\fpi\calI ,\quad
\calJdl=e^5\fpi^3\calJ ,
\end{equation}
are characterized by a single parameter $\beta=\mpi/(e \fpi)$.
For a given value of $\beta$, we first solve the differential
equation for $F$ and then those for $(A,B,C)$ to obtain
$\bigl(\Mcldl,\calIdl,\calJdl\bigr)$.
Next, by identifying the Hamiltonian \eqref{H} for the nucleon
($\bm{I}^2=3/4$) and $\Delta$ ($\bm{I}^2=15/4$) with the inputs $M_N$
and $M_\Delta$, respectively, we get the corresponding
$(\fpi,e,\mpi)$. Namely, we solve
\begin{align}
M&=\frac{\fpi}{e}\left(\Mcldl+\frac12 e^4\calIdl\bmOmgdl^2
+\frac34 e^8\calJdl\bmOmgdl^4\right) ,
\\
\bm{I}^2&=\left(\calIdl+ e^4\calJdl\bmOmgdl^2\right)^2\bmOmgdl^2,
\label{I^2=()^2Omg^2}
\end{align}
with dimensionless $\bmOmgdl^2$ defined by
\begin{equation}
\bmOmgdl^2=\frac{1}{e^6\fpi^2}\,\bmOmg^2 ,
\label{bmOmgdl}
\end{equation}
for both the nucleon and $\Delta$
to obtain $(\fpi,e)$ and $\mpi=e\fpi\beta$.
We tune $\beta$ to reproduce the experimental value of $\mpi$.
Our result obtained this way is
\begin{equation}
\fpi=125\,\MeV , \quad e=5.64 ,
\label{our_fpi_e}
\end{equation}
which is realized at $\beta=0.196$.
Compared with the experimental value $\fpi^{\text{(exp)}}=186\,\MeV$,
our $\fpi$ is fairly improved from that of \cite{AN},
$\fpi^{\text{(AN)}}=108\,\MeV$, without relativistic correction.
The dimensionless quantities corresponding to the result
\eqref{our_fpi_e} are given by
\begin{equation}
\bigl(\Mcldl,\calIdl,\calJdl\bigr)=(37.9,70.2,279) ,
\end{equation}
and the angular velocities of the nucleon and $\Delta$ are as follows:
\begin{equation}
\abs{\bmOmg_N}=206\,\MeV
\quad(\bmOmgdl_N^2=8.44\times 10^{-5}),\qquad
\abs{\bmOmg_\Delta}=330\,\MeV
\quad(\bmOmgdl_\Delta^2=2.17\times 10^{-4}).
\label{valueOmegas}
\end{equation}
Note that the angular velocity of the nucleon ($\Delta$) has become
larger (smaller) than that in the rigid body approximation
(see eq.\ \eqref{valuesRB}).

In the rest of this subsection, we present graphically various
numerical results corresponding to $(\fpi,e)$ of \eqref{our_fpi_e}.

\noindent
\underline{\em Profiles of $(A,B,C)$}

\begin{figure}[htbp]
\begin{minipage}[t]{0.47\hsize}
\includegraphics[scale=0.7]{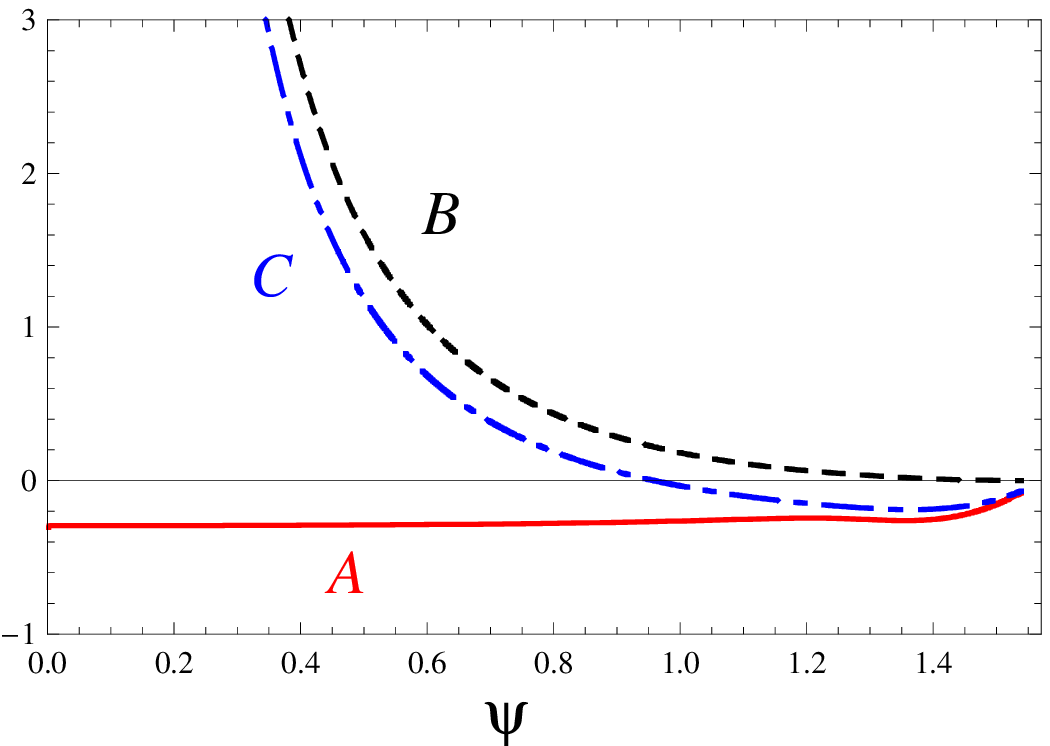}
\caption{\footnotesize{
Profiles of $A$ (red solid curve), $B$ (black broken one) and $C$
(blue dash-dotted one) at $\beta=0.196$. The horizontal coordinate is
$\psi=\tan^{-1}\rdl$.
As $r\to 0$ ($\psi\to 0$), while $A$ tends to a constant, $B$ and $C$
diverge as $O(1/r^2)$. As $r\to\infty$ ($\psi\to\pi/2$), all of them
tend to zero.
}}
\label{ABC_rev}
\end{minipage}
\hspace{5mm}
\begin{minipage}[t]{0.47\hsize}
\includegraphics[scale=0.7]{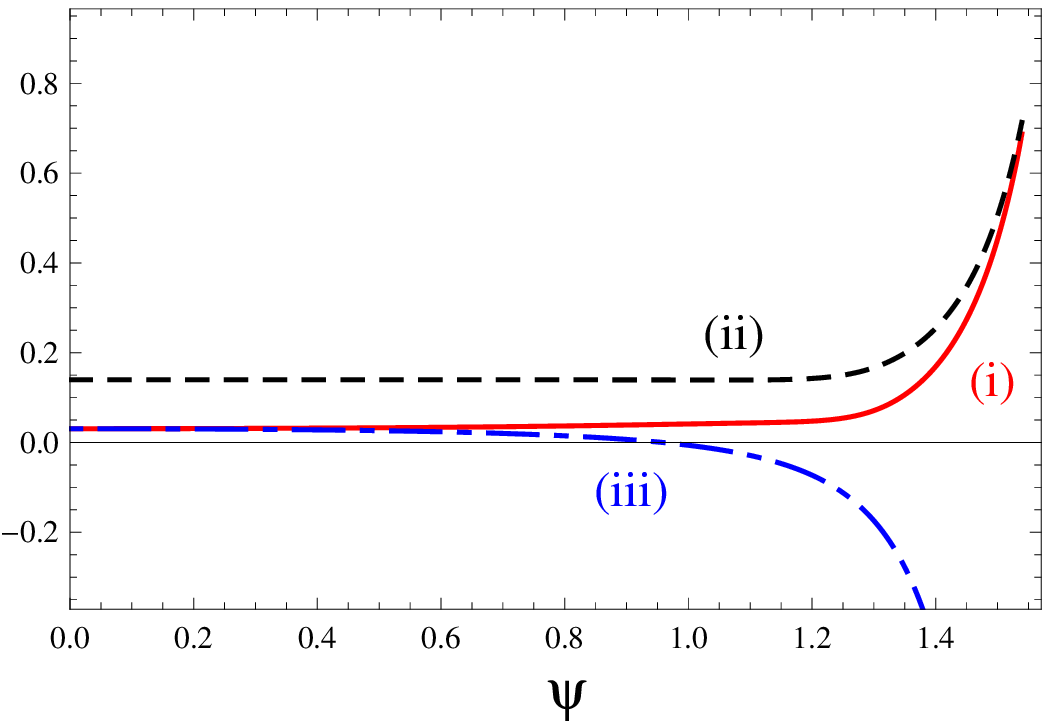}
\caption{\footnotesize{
Profiles of (i) $\bmOmg_N^2r^2(C-A)$ (red solid curve),
(ii) $\bmOmg_N^2r^2 Y$ (black broken one), and
(iii) $\bmOmg_N^2r^2 C$ (blue dash-dotted one) at $\beta=0.196$
as functions of $\psi=\tan^{-1}\rdl$.
As $r\to 0$, the three functions all approach constants.
As $r\to\infty$, while $\bmOmg_N^2r^2(C-A)$ and $\bmOmg_N^2r^2 Y$
go to the same constant, $\bmOmg_N^2/(2\mpi^2)=1.11$
(c.f., \eqref{DE_ABC_infty_series}),
$\bmOmg_N^2r^2 C$ diverges linearly in $r$.
}}
\label{r2ABC_rev}
\end{minipage}
\end{figure}

First, in Figs.\ \ref{ABC_rev} and \ref{r2ABC_rev}, the profiles of
$(A,B,C)$ as functions of $\psi=\tan^{-1}e\fpi r$ ($0\le\psi<\pi/2$)
are given.
Fig.\ \ref{ABC_rev} shows the profiles of $(A,B,C)$ themselves, and
Fig.\ \ref{r2ABC_rev} shows those of $\bmOmg^2r^2(C-A,Y,C)$, which are
functions appearing in \eqref{uinp} and are directly relevant to the
deformation of baryons. As $\bmOmg^2$ we have taken the nucleon
angular velocity $\bmOmg_N^2=(206\,\MeV)^2$.

\noindent
\underline{\em Profiles of the integrands of $(\Mcl,\calI,\calJ)$}

\begin{figure}[htbp]
\begin{minipage}[t]{0.47\hsize}
\includegraphics[scale=0.7]{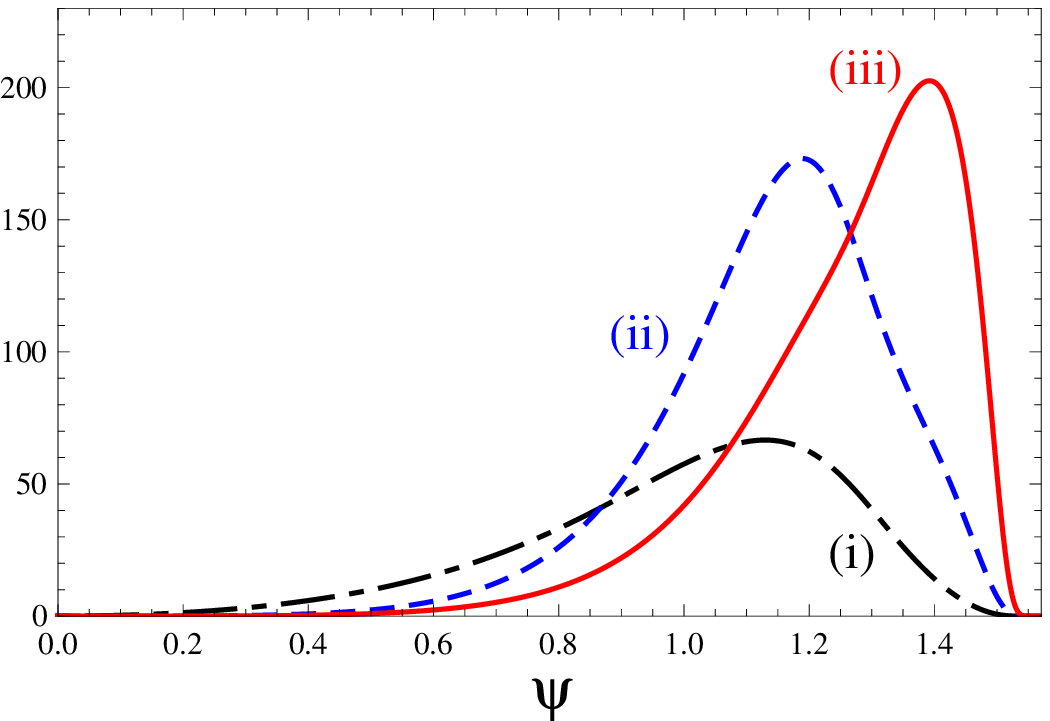}
\caption{\footnotesize{
Profiles of the integrands of $\psi$-integrations for
(i) $\Mcldl$ (black dash-dotted curve),
(ii) $\calIdl$ (blue broken one), and
(iii) $(1/4)\calJdl$ (red solid one), at $\beta=0.196$.
Note that the profile for $\calJdl$ is multiplied by $(1/4)$.
The horizontal coordinate is $\psi=\tan^{-1}\rho$.
}}
\label{theMIJ}
\end{minipage}
\hspace{5mm}
\begin{minipage}[t]{0.47\hsize}
    \includegraphics[scale=0.7]{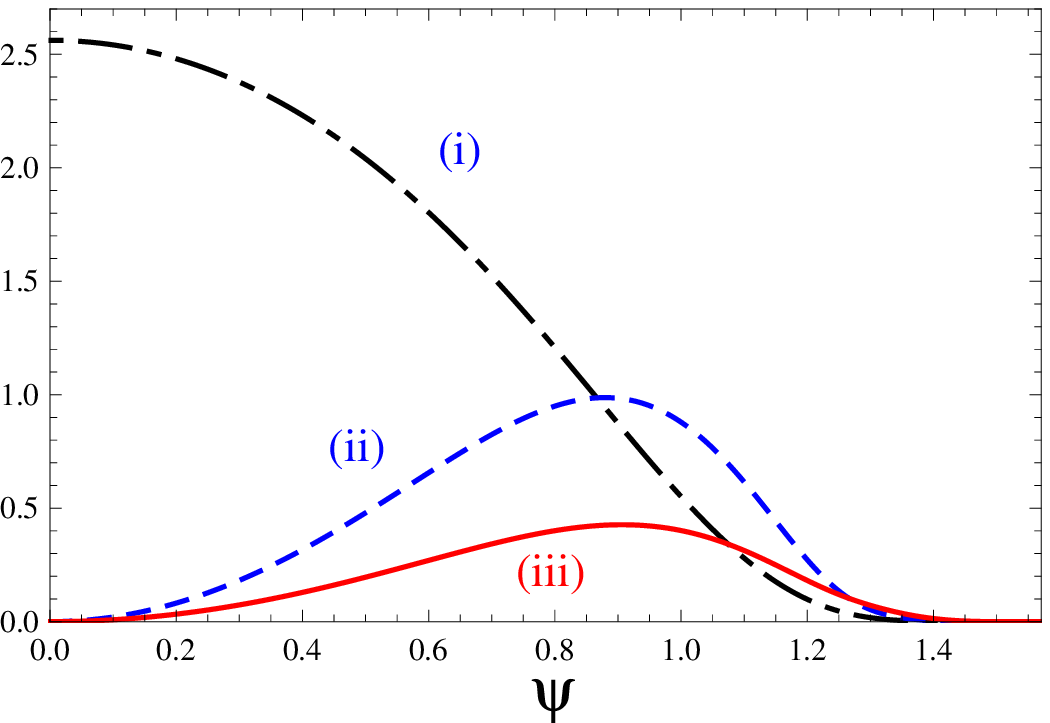}
\caption{\footnotesize{
Profiles of the angle averaged densities in real space of
(i) $\Mcldl$ (black dash-dotted curve),
(ii) $\calIdl$ (blue broken one), and
(iii) $(1/4)\calJdl$ (red solid one), at $\beta=0.196$.
The horizontal coordinate is $\psi=\tan^{-1}\rho$.
}}
\label{laMIJ}
\end{minipage}
\end{figure}

Next, Figs.\ \ref{theMIJ} and \ref{laMIJ} show the profiles of the
integrands of the dimensionless quantities
$\bigl(\Mcldl,\calIdl,\calJdl\bigr)$ constituting the Hamiltonian.
They are given as $\rdl$-integrations; \eqref{Mcl}, \eqref{calI} and
\eqref{calJ} with the replacements $r\to\rdl$, $d/dr\to d/d\rdl$,
$\mpi\to\beta$ and $(\fpi,e)\to(1,1)$.
In Fig.\ \ref{theMIJ}, we show the profiles of their integrands
as the $\psi$-integrations. Namely, giving $\calO=\Mcldl$,
$\calIdl$, $(1/4)\calJdl$ as
$\calO=\int_0^{\pi/2}\!d\psi\,f_\calO(\psi)$,
Fig.\ \ref{theMIJ} shows the profiles of the three $f_\calO(\psi)$
(for the sake of visualization, we take $(1/4)\calJdl$ instead
of $\calJdl$ itself).
On the other hand, expressing $\calO$ differently as
$\calO=4\pi\int_0^{\pi/2}\!d\rdl\,\rdl^2\,g_\calO(\psi)$,
Fig.\ \ref{laMIJ} shows the profiles of $g_\calO(\psi)$, which
are the angle averaged densities in real space.
We have $f_\calO(\psi)=4\pi(\sin^2\psi/\cos^4\psi)\,g_\calO(\psi)$.
\begin{figure}[htbp]
\begin{minipage}[t]{0.47\hsize}
\includegraphics[scale=0.7]{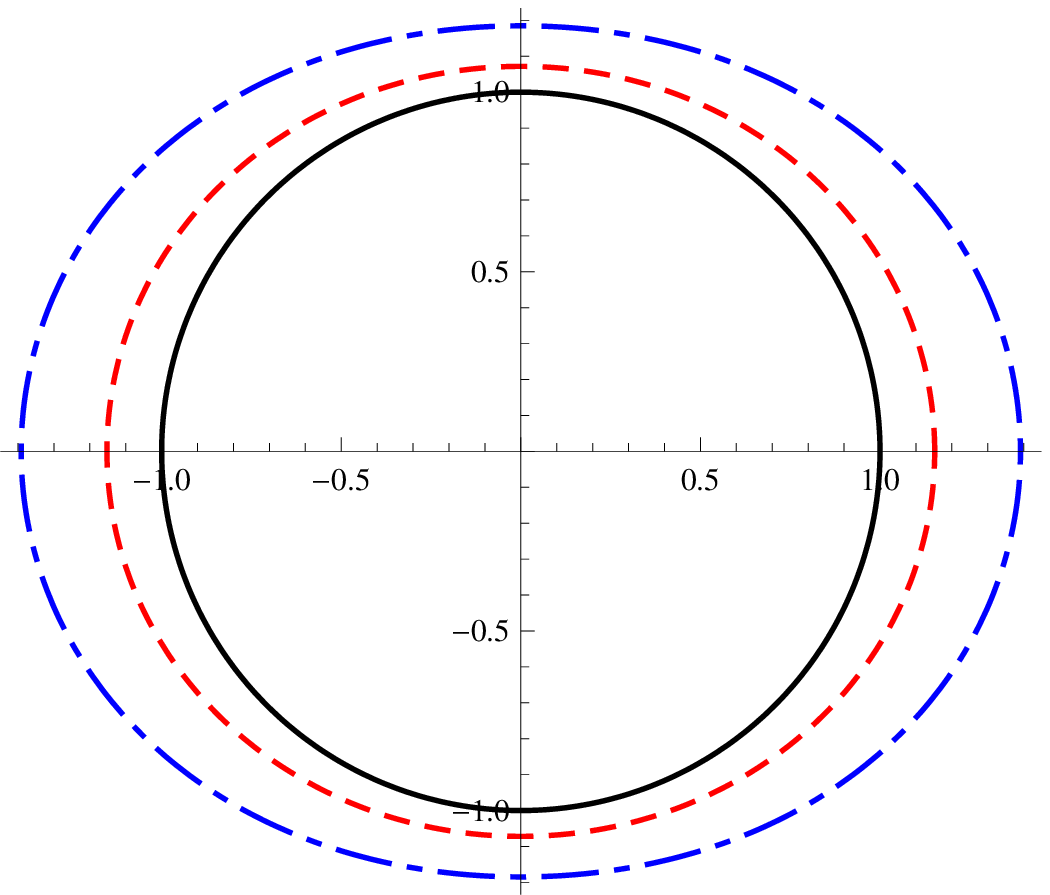}
\caption{\footnotesize{
The spheroids \eqref{riny} of the nucleon (red broken ellipse curve)
and $\Delta$ (blue dash-dotted one) spinning around the vertical axis
and corresponding to the original
sphere (black solid circle) with radius $a=1\,\fm$.
The units on the horizontal and vertical axes are $\fm$.
}}
\label{deformation:circle}
\end{minipage}
\hspace{5mm}
\begin{minipage}[t]{0.47\hsize}
\includegraphics[scale=1.15]{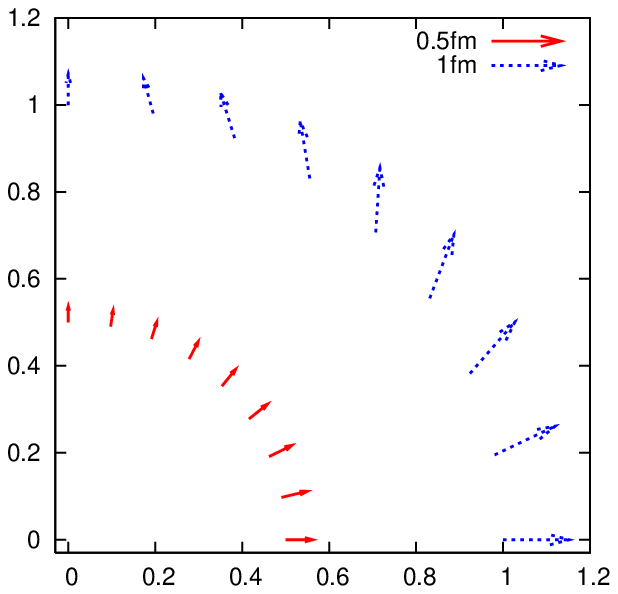}
\caption{\footnotesize{
Deformation vectors $-\bm{u}$ \eqref{uinp} of the nucleon at various
angles $\theta$.
The red solid arrows and blue broken ones are $-\bm{u}$ on
$r=0.5\,\fm$ and $r=1\,\fm$, respectively.
The units on the horizontal and vertical axes are $\fm$.
}}
\label{deformation:field}
\end{minipage}
\end{figure}

\noindent
\underline{\em Deformation of baryons}

As we explained in Sec.\ \ref{sec:ISSF}, the spinning Skyrmion field
\eqref{OSS} describes a spheroid \eqref{riny}.
In Fig.\ \ref{deformation:circle}, we show the spheroids (ellipses)
which represent the nucleon and $\Delta$ and correspond to the sphere
of radius $a=1\,\fm$ without spinning motion.
Namely, they are the spheroids \eqref{riny} with $\bmOmg^2$ given by
$\bmOmg_N^2$ and $\bmOmg_\Delta^2$ of \eqref{valueOmegas},
respectively, and with $a=1\,\fm$.
In Fig.\ \ref{deformation:field}, we show the deformation vectors
$-\bm{u}$ \eqref{uinp} of the nucleon representing the shift of each
point on the original sphere due to the spinning motion at radii
$0.5\,\fm$ and $1\,\fm$.
Note that $\bmy$ is on the sphere and $\zbmy=\bmy-\bm{u}$ is on the
spheroid. In both the figures, the angular velocity points to the
vertical direction.

\subsection{Static properties of nucleons}
\label{sec:SPN}

Having determined the parameters $(\fpi,e)$ of the Skyrme model from
the experimental values of $(M_N,M_\Delta,\mpi)$, let us calculate the
various static properties of nucleons which were analyzed in the rigid
body approximation in Refs.\ \cite{ANW,AN}.
Concretely, we consider the charge radii, magnetic moments,
magnetic charge radii, and axial vector coupling of nucleons.
We are of course interested in how the relativistic correction
modifies the numerical results from those of \cite{AN}.

First, we need the analytic expressions of these static
properties derived from the Lagrangian density \eqref{calL_Skyrme} of
the Skyrme model and the spinning Skyrmion field \eqref{OSS}.
In Appendix \ref{app:AE}, we present the definitions, outline of
derivations and the final analytic expressions of the static
properties. The analytic expressions are as follows:
\begin{itemize}
\item
Isoscalar mean square charge radius $\VEV{r^2}_{I=0}$:
eq.\ \eqref{ISCR}.

\item
Isovector mean square charge radius $\VEV{r^2}_{I=1}$:
eq.\ \eqref{IVCR}.

\item
Isoscalar mean square magnetic radius $\VEV{r^2}_{M,I=0}$:
eq.\ \eqref{ISMCR}.

\item
Isovector mean square magnetic radius $\VEV{r^2}_{M,I=1}$:
eq.\ \eqref{IVMCR}.
This is in fact equal to the electric one $\VEV{r^2}_{I=1}$.

\item
Isoscalar $g$-factor $g_{I=0}$: eq.\ \eqref{g_I=0}.

\item
Isovector $g$-factor $g_{I=1}$: eq.\ \eqref{g_I=1} together with
\eqref{wtg_I=1}.
It has another and simpler expression \eqref{g_I=1_simpler}.

\item
Axial vector coupling constant $g_A$: eq.\ \eqref{g_A} together with
\eqref{wtg_A^(1)} and \eqref{wtg_A^(3)}.
\end{itemize}
These quantities are all given as integrations over the
dimensionless radial coordinate $\rdl=e\fpi r$. The mean square charge
radii are given in units of $1/\fpiesq=(0.280\,\fm)^2$.
In every quantity, the relativistic correction part is multiplied by
$e^4\bmOmgdl_N^2$ with $\bmOmgdl_N^2$ being the dimensionless angular
velocity of the nucleon \eqref{valueOmegas}.

In Table \ref{tbl:staticP}, we summarize the numerical values of the
static properties obtained by using $(e,\fpi,\bmOmgdl_N^2)$ and the
functions $F$ and $(A,B,C)$ determined in Sec.\ \ref{sec:DetfpieABC}.
Instead of the $g$-factors, we present the nucleon magnetic moments
$\mu_p$ and $\mu_n$ in units of Bohr magneton:
\begin{equation}
\mu_p=\frac14\left(g_{I=0}+g_{I=1}\right),
\qquad
\mu_n=\frac14\left(g_{I=0}-g_{I=1}\right) .
\end{equation}
In the table, we also present the predictions of Ref.\ \cite{AN}
without relativistic correction as well as the experimental values.

\begin{table}[htbp]
\begin{center}
\begin{tabular}{c|cccc}
\toprule
&\parbox{23mm}{Prediction \\(this paper)}
&\parbox{23mm}{Prediction \\(Ref.\ \cite{AN})}
&Experiment
\\ \hline
$\fpi$ & $125\,\MeV$ & $108\,\MeV$ & $186\,\MeV$
\\ \hline
$\VEV{r^2}_{I=0}^{1/2}$& $0.59\,\fm$ & $0.68\,\fm$ & $0.81\,\fm$
\\
$\VEV{r^2}_{I=1}^{1/2}$& $1.17\,\fm$ & $1.04\,\fm$ & $0.94\,\fm$
\\
$\VEV{r^2}_{M,I=0}^{1/2}$& $0.85\,\fm$ &$0.95\,\fm$ & $0.82\,\fm$
\\
$\VEV{r^2}_{M,I=1}^{1/2}$& $1.17\,\fm$ & $1.04\,\fm$& $0.86\,\fm$
\\
$\mu_p$ & $1.65$ & $1.97$ & $2.79$
\\
$\mu_n$ & $-0.99$ & $-1.24$ & $-1.91$
\\
$|\mu_p/\mu_n|$ & $1.67$ & $1.59$ & $1.46$
\\
$g_A$ & $0.58$ & $0.65$ & $1.24$ \\
\bottomrule
\end{tabular}
\caption{
The static properties of nucleons.
Prediction of this paper and that of Ref.\ \cite{AN} in the rigid body
approximation both use the experimental values of
$(M_N,M_\Delta,\mpi)$ as inputs.
We follow the notations of Ref.\ \cite{ANW}.}
\label{tbl:staticP}
\end{center}
\end{table}

As seen from the table, the difference between the prediction of this
paper and that of Ref.\ \cite{AN} is in the range of $5\%$ to $20\%$.
(Note that each of our predictions is not given simply by adding
the $\bmOmg^2$ correction to the result of Ref.\ \cite{AN} since the
values of $\fpi$ and $e$ are also changed.)
Although the fundamental parameter of the theory, $\fpi$, has been
improved, the relativistic correction makes the value of theoretical
prediction further away from the experimental one
for most of the static properties of nucleons (the only exception is
$\VEV{r^2}_{M,I=0}$).
However, this should not be regarded as a manifestation of problems
of our basic EOM principle for introducing collective coordinates.
There would be two possibilities for this unwelcome result.
One is that, since the relativistic correction is rather large, we
have to take into account the contributions from still higher orders
in $\bmOmg^2$ for obtaining better results of the static properties of
nucleons.
Another possibility would be that we cannot expect the Skyrme theory,
which is merely a low energy effective theory, to reproduce precisely
the physics of the baryon sector even if the full relativistic
treatment of rotational collective coordinate is carried out.
Concerning the first possibility, we should note that the ratio of the
contributions of the three terms of the Hamiltonian \eqref{H} to the
baryon masses is $89:7:4$ for the nucleon and $68:14:18$ for
$\Delta$.
This suggests us that the expansion in powers of the angular velocity
$\bmOmg$ is not a good approximation especially for $\Delta$ with
a larger (iso)spin.

\section{Summary and outlook}
\label{sec:SD}

In this paper, we proposed the EOM principle for constructing the
relativistic system of collective coordinates of a field theory
soliton, and applied it to the ``next to the rigid body approximation''
analysis of the spinning Skyrmion.
The relativistic correction was successfully incorporated in the
argument of the static Skyrmion solution in terms of the three
functions $(A(r),B(r),C(r))$ obeying a consistent set of
differential equations demanded by the EOM principle.
We computed the decay constant and various static properties of
nucleons with leading relativistic corrections by taking the masses of
the nucleon, $\Delta$ and the pion as inputs.
The relativistic corrections to the results in the rigid body
approximation are found to be in the range of $5\%$ -- $20\%$.
Though the value of the decay constant has
become closer to the experimental one due to the correction, the
results are not good for most of the static properties of nucleons.
We also studied how the baryons deform from the spherical shape to the
spheroidal one due to the spinning motion.

Since we have introduced the relativistic corrections in a proper
manner based on the EOM principle, our unwelcome result on the static
properties of nucleons is an inevitable nature of the baryon sector of
the Skyrme model: It should not be regarded as a problem of our basic
EOM principle.
If the simple Skyrme model can reproduce the real baryon physics more
precisely, a remedy for the unwelcome results should be looked for
in our use of expansion in powers of the angular velocity.
Though we took into account only the first relativistic correction of
$O(\bmOmg^4)$ in this paper, it would be necessary to consider
higher order corrections, or to devise an approximation valid for an
extreme large angular velocity, at least for analyzing $\Delta$.
For this, we have to consider a systematic generalization of the
expressions \eqref{absbmy=} and \eqref{tan=(...)tan} in terms of the
Legendre polynomials.
The analysis \cite{Piette:1994mh} of the spinning baby Skyrmion on the
two-dimensional plane may be helpful for considering the extreme
case in the $3+1$ dimensional model.
With a better (or the full) realization of the EOM principle, we can
give more reliable predictions for the Skyrmion.
It is also an important subject to apply our method to the collective
coordinate quantization of other interesting physical systems.

\section*{Acknowledgements}
We would like to thank Tatsuo Azeyanagi, Mitsutoshi Fujita, Kenji
Fukushima, Koji Hashimoto, Koh Iwasaki, Antal Jevicki, Hikaru Kawai,
Shin Nakamura, Keisuke Ohashi, Kentaroh Yoshida and Koichi Yoshioka
for valuable discussions.
We would also like to thank organizers and participants of workshops
"KEK Theory Workshop 2010" at KEK and "Solitons and QCD" at RIKEN for
helpful conversations.
Discussions during the YITP workshop YITP-W-08-04 on ``Development of
Quantum Field Theory and String Theory'' were useful in completing
this work.
The work of H.~H.\ was supported in part by a Grant-in-Aid for
Scientific Research (C) No.~21540264 from the Japan Society for the
Promotion of Science (JSPS).  The work of T.~K.\ was supported by
a Grand-in-Aid for JSPS Fellows No.~21-951.
The numerical calculations were carried out on Altix3700 BX2 at YITP
in Kyoto University.  This work was supported by the Grant-in-Aid for
the Global COE Program "The Next Generation of Physics, Spun from
Universality and Emergence" from the Ministry of Education, Culture,
Sports, Science and Technology (MEXT) of Japan.

\appendix

\section{Derivation of eq.\ (\ref{FEOM_OSS})}
\label{app:deriv_FEOM_OSS}

In this appendix, we sketch the derivation of \eqref{FEOM_OSS},
namely, the LHS of the Skyrme field theory EOM \eqref{FEOM} with the
improved spinning Skyrmion field \eqref{OSS} substituted.
In this derivation, the most important formula is \eqref{LmubyLcl}
for $L_\mu(\bmx,t)$ with $\Lcl_a(\bmy)$ $(a=1,2,3)$ given by
\begin{align}
\Lcl_a(\bmy)&=-i\Ucl(\bmy)\Drv{}{y_a}\Ucl(\bmy)^\dagger
\notag\\
&=-\frac{1}{2r}\Bigl\{\sin 2F\delta_{ab}
+\left(2rF'-\sin 2F\right)\why_a\why_b
-\left(1-\cos 2F\right)\epsilon_{abc}\why_c\Bigr\}\tau_b ,
\label{Lcl_a=}
\end{align}
where $\why_a$ is a unit vector $\why_a=y_a/\abs{\bmy}$.
In this appendix, $r$ denotes the length of $\bmy$, and the argument
$r$ of the function $F$ and its derivatives is omitted:
\begin{equation}
r=\abs{\bmy},\qquad F= F(r),\qquad F'=\drv{F(r)}{r}.
\end{equation}
As for the argument $r$ of $(A,B,C)$, the difference between
$r=\abs{\bmx}$ and $r=\abs{\bmy}$ does not matter since $(A,B,C)$ are
already multiplied by the $O(\p_t^2)$ quantities in \eqref{bmy=}.

\noindent
\underline{\em Necessary formulas}

Let us present the formulas of the differentiations of $\bmy(\bmx,t)$
\eqref{bmy=} with respect to $\bmx$ and $t$. First, the
time-derivatives are given by
\begin{align}
\Drv{y_a(\bmx,t)}{t}
&=-(\RiRd\bmy)_a+2\left[A(\RiRd\bmy)_b(\dtRiRd\bmy)_b
+r^2 B\Tr(\dtRiRd\RiRd)\right]y_a
\notag\\
&\qquad\qquad
+r^2 C\bigl(\bigl\{\dtRiRd,\RiRd\bigr\}\bmy\bigr)_a
+O(\p_t^5) ,
\label{py/pt}
\\
\Drv{^2y_a(\bmx,t)}{t^2}&=-(\dtRiRd\bmy)_a+\bigl((\RiRd)^2\bmy\bigr)_a
+O(\p_t^4) ,
\label{p^2y_a/pt^2}
\end{align}
where $\dtRiRd$ is defined by
\begin{equation}
\dtRiRd=\drv{}{t}\RiRd .
\label{dtRiRd}
\end{equation}
Note that, in \eqref{py/pt} and \eqref{p^2y_a/pt^2}, we have
presented their expressions in terms of $\bmy$, which are obtained by
rewriting the original ones in terms of $\zbmy$ by using \eqref{bmy=}.
The counting of the order of time-derivatives $\p_t$ is of course is
defined by regarding $\bmx$ (and not $\bmy$) as a zero-th order
quantity.
Though we need only the first term on the RHS of \eqref{py/pt} in
this appendix, we have presented the expression valid to $O(\p_t^4)$.
This will be used in deriving the coefficient $\calJ$ of \eqref{calJ}
in Appendix \ref{app:calc_calJ}.

Next, for the $\bmx$-derivatives, we have
\begin{align}
\Drv{y_a(\bmx,t)}{x_i}&=R_{ib}\left(\delta_{ab}+\yx{ab}\right) ,
\label{py/px}
\\
\Drv{^2 y_a(\bmx,t)}{x_i\p x_j}&=R_{ib}R_{jc}\yxx{a}{bc} ,
\label{p^2y/pxpx}
\end{align}
where $\yx{ab}$ and $\yxx{a}{bc}$ are given by
\begin{align}
\yx{ab}&=A\left[(\RiRd\bmy)^2\delta_{ab}
-2 y_a\bigl((\RiRd)^2\bmy\bigr)_b\right]
+\frac{1}{r}\drv{A}{r}(\RiRd\bmy)^2y_a y_b
\notag\\
&\quad
+\Tr(\RiRd)^2\left[r^2 B\,\delta_{ab}
+\left(r\drv{B}{r}+2B\right)y_a y_b\right]
\notag\\
&\quad
+\left(2C +r\drv{C}{r}\right)\bigl((\RiRd)^2\bmy\bigr)_a y_b
+r^2 C\bigl((\RiRd)^2\bigr)_{ab}+O(\p_t^4) ,
\label{yx}
\end{align}
and
\begin{align}
&\yxx{a}{bc}=\yxx{a}{cb}\equiv \Drv{\yx{ab}}{y_c}
=-2A\left[\delta_{ab}\bigl((\RiRd)^2\bmy\bigr)_c
+\delta_{ac}\bigl((\RiRd)^2\bmy\bigr)_b
+y_a\bigl((\RiRd)^2\bigr)_{bc}\right]
\notag\\
&
+\frac{1}{r}\drv{A}{r}\biggl[
(\RiRd\bmy)^2\Bigl(\delta_{ab}y_c+\delta_{ac}y_b
+y_a\bigl(\delta_{bc}-\wh{y}_b\wh{y}_c\bigr)\Bigr)
\!-\! 2y_a\!\left(\bigl((\RiRd)^2\bmy\bigr)_by_c
+\bigl((\RiRd)^2\bmy\bigr)_cy_b\right)
\biggr]
\notag\\
&+\drv{^2A}{r^2}\,(\RiRd\bmy)^2\,y_a\wh{y}_b\wh{y}_c
\notag\\
&+\left(2B+r\drv{B}{r}\right)\Tr(\RiRd)^2\bigl(
\delta_{ab}y_c+\delta_{ac}y_b+y_a\delta_{bc}\bigr)
+\left(\frac{3}{r}\drv{B}{r}+\drv{^2 B}{r^2}\right)
\Tr(\RiRd)^2\,y_ay_by_c
\notag\\
&+\left(2C+r\drv{C}{r}\right)\left[\bigl((\RiRd)^2\bigr)_{ab}y_c
+\bigl((\RiRd)^2\bigr)_{ac}y_b
+\bigl((\RiRd)^2 \bmy\bigr)_a\delta_{bc}
\right]
\notag\\
&+\left(\frac{3}{r}\drv{C}{r}+\drv{^2C}{r^2}\right)
\bigl((\RiRd)^2\bmy\bigr)_ay_by_c  +O(\p_t^4).
\label{yxx}
\end{align}
In particular, we have
\begin{align}
\yx{ab}+\yx{ba}
&=2\left[A\,(\RiRd\bmy)^2+r^2 B\Tr(\RiRd)^2\right]
\delta_{ab}
\notag\\
&\quad
+2\left[\frac{1}{r}\drv{A}{r}(\RiRd\bmy)^2
+\left(r\drv{B}{r}+2B\right)\Tr(\RiRd)^2\right]y_a y_b
\notag\\
&\quad
+\left(-2A+2C+r\drv{C}{r}\right)
\left[\bigl((\RiRd)^2\bmy\bigr)_a y_b
+\bigl((\RiRd)^2\bmy\bigr)_b y_a\right]
+2r^2 C\bigl((\RiRd)^2\bigr)_{ab} ,
\label{yx_ab+yx_ba}
\end{align}
and
\begin{align}
\yxx{a}{bb}
&=\left(\drv{^2 A}{r^2}+\frac{8}{r}\drv{A}{r}\right)
(\RiRd\bm{y})^2\,y_a
+\left(r^2\drv{^2 B}{r^2}+8r\drv{B}{r}-2A+10B\right)
\Tr(\RiRd)^2\,y_a
\notag\\
&\quad
+\left(r^2\drv{^2 C}{r^2}+8r\drv{C}{r}
-4A+10 C\right)\bigl((\RiRd)^2\bm{y}\bigr)_a .
\end{align}

We also need formulas concerning $\Lcl_a(\bmy)$. First, we have
\begin{align}
\Drv{\Lcl_b(\bmy)}{y_a}
&=\frac{1}{r^2}\Biggl\{
\left[-r^2 F''+\left(2+\cos 2F\right)r F'
-\frac32\sin 2F\right]\why_a\why_b\why_c
\notag\\
&\qquad
+\left(\frac12\sin 2F-\cos 2F\,r F'\right)\why_a\delta_{bc}
+\left(\frac12\sin 2F -r F'\right)
\left(\why_b\delta_{ac}+\why_c\delta_{ab}\right)
\notag\\
&\qquad
+\bigl(1-\cos 2F-\sin 2F\,r F'\bigr)\why_a\epsilon_{bdc}\why_d
+\frac12\left(1-\cos 2F\right)\epsilon_{abc}
\Biggr\}\tau_c .
\end{align}
Next, the double commutator of $\Lcl_a(\bmy)$ is given by
\begin{align}
&T_{abc}(\bmy)=-T_{acb}(\bmy)
\equiv[\Lcl_a(\bmy),[\Lcl_b(\bmy),\Lcl_c(\bmy)]\bigr]
\notag\\
&=-\frac{1}{r}\left[2(F')^2-\frac{1-\cos 2F}{r^2}\right]
\why_a\Bigl\{\sin 2F\,
\left(\wh{y}_b\delta_{cd}-\wh{y}_c\delta_{bd}\right)
+(1-\cos 2F)\left(
\wh{y}_b\epsilon_{ced} -\wh{y}_c\epsilon_{bed}\right)\wh{y}_e
\Bigr\}\tau_d
\notag\\
&\quad
-\frac{1-\cos 2F}{r^3}\biggl\{
\sin 2F\left(\delta_{ab}\delta_{cd}-\delta_{ac}\delta_{bd}\right)
+\left(2rF'-\sin 2F\right)\left(
\delta_{ab}\wh{y}_c-\delta_{ac}\wh{y}_b\right)\wh{y}_d
\notag\\
&\qquad\qquad\qquad\qquad
+(1-\cos 2F)\left(\delta_{ab}\epsilon_{ced}
-\delta_{ac}\epsilon_{bed}\right)\wh{y}_e
\biggr\}\tau_d .
\label{T_abc}
\end{align}
Upon contraction, we have
\begin{align}
T_{bab}(\bmy)&=\left[\frac{2}{r}\left(F'\right)^2
+\frac{1-\cos 2F}{r^3}\right]
\Bigl(\sin 2F\,\delta_{ab}
+\left(1-\cos 2F\right)\epsilon_{acb}\wh{y}_c\Bigr)\tau_b
\notag\\
&\quad
-\left[2\,\frac{\sin 2F}{r}\left(F'\right)^2
-4\frac{1-\cos 2F}{r^2}\,F'
+\frac{\sin 2F(1-\cos 2F)}{r^3}\right]\wh{y}_a\wh{y}_b\tau_b .
\label{T_bab}
\end{align}

Finally, the orthogonality,
\begin{equation}
(\RiRd\bmy)_a y_a=0,
\label{Orthg}
\end{equation}
which is due to the anti-symmetric nature of the matrix $\RiRd$, often
simplifies the calculations in many places.

\noindent
\underline{\em Evaluation of each term in \eqref{FEOM}}

Let us obtain the expression of each term in \eqref{FEOM} in terms
of $\Lcl_a(\bmy)$ and its derivative. We keep only terms containing at
most two time-derivatives.
First, for $\p_\mu L^\mu=-\p_0 L_0+\p_i L_i$, we have\begin{align}
\p_0L_0(\bmx,t)&=\Drv{^2y_a}{t^2}\Lcl_a(\bmy)
+\Drv{y_a}{t}\Drv{y_b}{t}(\p_a\Lcl_b)(\bmy)
\label{p_0L_0}
\notag\\
&=\left[-(\dtRiRd\bmy)_a+\bigl((\RiRd)^2\bmy\bigr)_a\right]
\Lcl_a(\bmy)
+(\RiRd\bmy)_a(\RiRd\bmy)_b(\p_a\Lcl_b)(\bmy) ,
\\
\p_i L_i(\bmx,t)&=\Drv{^2 y_a}{x_i^2}\Lcl_a(\bmy)
+\Drv{y_a}{x_i}\Drv{y_b}{x_i}(\p_a\Lcl_b)(\bmy)
\notag\\
&=(\p_a\Lcl_a)(\bmy)+\yxx{a}{bb}\Lcl_a(\bmy)
+\left(\yx{ab}+\yx{ba}\right)(\p_a\Lcl_b)(\bmy) ,
\end{align}
with $(\p_a\Lcl_b)(\bmy)=(\p/\p y_a)\Lcl_b(\bmy)$.

Next, let us consider the part of EOM coming from the Skyrme term:
\begin{equation}
\p_\mu\bigl[L_\nu,[L^\mu,L^\nu]\bigr]
=-\p_0\bigl[L_i,[L_0,L_i]\bigr]-\p_i\bigl[L_0,[L_i,L_0]\bigr]
+\p_i\bigl[L_j,[L_i,L_j]\bigr] .
\label{FEOM_Skyrm}
\end{equation}
For the first two terms on the RHS, we can substitute the lowest order
expressions $L_0=-(\RiRd\bmy)_a\Lcl_a(\bmy)$ and
$L_i=R_{ia}\Lcl_a(\bmy)$ to get
\begin{align}
\p_0\bigl[L_i,[L_0,L_i]\bigr]
&=\left[-(\dtRiRd\bmy)_a+\bigl((\RiRd)^2\bmy\bigr)_a\right]
T_{bab}(\bmy)
+(\RiRd\bmy)_a(\RiRd\bmy)_c\Drv{}{y_c}T_{bab}(\bmy) ,
\\
\p_i\bigl[L_0,[L_i,L_0]\bigr]&=\Drv{}{y_b}\left(
(\RiRd\bmy)_a(\RiRd\bmy)_c T_{abc}(\bmy)\right) ,
\end{align}
with $\dtRiRd$ defined by \eqref{dtRiRd}.
The last term \eqref{FEOM_Skyrm} is given, using \eqref{py/px} and
\eqref{p^2y/pxpx}, by
\begin{align}
\p_i\bigl[L_j,[L_i,L_j]\bigr]
&=\Drv{}{x_i}\left(\Drv{y_a}{x_j}\Drv{y_b}{x_i}\Drv{y_c}{x_j}
\,T_{abc}(\bmy)\right)
\notag\\
&=\Drv{}{y_a}T_{bab}+\yxx{a}{cc}T_{bab}+\yxx{c}{ab}T_{abc}
+\left(\Upsilon_{ac}+\Upsilon_{ca}\right)\left(
\Drv{}{y_c}T_{bab}+\Drv{}{y_b}T_{abc}\right) ,
\label{p_i[L_j,[L_i,L_j]]}
\end{align}
where we have used $\yxx{a}{bc}T_{abc}=0$.

\noindent
\underline{\em Total of the LHS of \eqref{FEOM}}

{}From the above results, \eqref{p_0L_0}--\eqref{p_i[L_j,[L_i,L_j]]}, we
finally obtain
\begin{align}
&\mbox{LHS of \eqref{FEOM}}\bigr|_{U(\bmx,t)=\Ucl(\bmy)}
=\Drv{\calLcl_a(\bmy)}{y_a}
-i\mpi\left(\Ucl(\bmy)-\frac12\tr\Ucl(\bmy)\right)
\notag\\
&\quad
+\left((\dtRiRd\bmy)_a
-\bigl((\RiRd)^2\bmy\bigr)_a+\yxx{a}{bb}\right)\calLcl_a(\bmy)
+\left(-(\RiRd\bmy)_a(\RiRd\bmy)_b+\Upsilon_{ab}+\Upsilon_{ba}
\right)\Drv{\calLcl_a(\bmy)}{y_b}
\notag\\
&\quad
-\frac{1}{\fpiesq}\left\{\yxx{c}{ab}T_{abc}(\bmy)
+\left(\Upsilon_{ac}+\Upsilon_{ca}\right)\Drv{T_{abc}(\bmy)}{y_b}
-\Drv{}{y_b}\left[
(\RiRd\bmy)_a(\RiRd\bmy)_c T_{abc}(\bmy)\right]\right\} ,
\label{calc_FEOM_OSS}
\end{align}
with $\calLcl_a$ defined by
\begin{equation}
\calLcl_a=\Lcl_a -\frac{1}{\fpiesq}T_{bab} .
\end{equation}
The first two terms containing no time-derivatives on the RHS of
\eqref{calc_FEOM_OSS} vanish:
\begin{equation}
\Drv{\calLcl_a(\bmy)}{y_a}
-i\mpi\left(\Ucl(\bmy)-\frac12\tr\Ucl(\bmy)\right)=0 .
\end{equation}
This is nothing but the EOM of $\Ucl$.
The next term, $(\dtRiRd\bmy)_a\calLcl_a$, gives the first term on the
RHS of \eqref{FEOM_OSS}, and the remaining terms give the EOM-breaking
part containing $\EQ_n$ ($n=1,2,3,4$).
In the latter calculation, the orthogonality \eqref{Orthg} greatly
simplifies our lengthy task.

Finally, by dropping the terms containing $(A,B,C)$ in
\eqref{calc_FEOM_OSS}, namely, by removing $\yx{ab}$ and $\yxx{a}{bc}$,
we obtain \eqref{V_FEOM} before introducing the improvement.

\section{Expressions of $\bm{\EQ_{n=1,2,3,4}}$}
\label{app:EQ_1234}

In this appendix, we present concrete (and lengthy) expressions of the
four quantities $\EQ_n$ ($n=1,2,3,4$) appearing in \eqref{FEOM_OSS}.
We also present the approximate expressions of the differential
equations for $(A,C)$ and $Y$ near the infinity and their
solutions.
The primes on $F$ denote differentiations with respect to $r$.
First, $\EQ_n$ are given as follows:
\begin{align}
\EQ_1&=\frac{2}{r^2}F'A-F'\drv{^2B}{r^2}
-2\left(F''+\frac{4}{r}F'\right)\drv{B}{r}
-\frac{2}{r}\left(3F''+\frac{7}{r}F'-\frac{1}{r^2}\sin2F\right)B
\notag\\
&-\frac{1}{r^2}\left(2F'-\frac{1}{r}\sin2F\right)C
+\frac{1}{\fpiesq}\Biggl\{\frac{4}{r^4}(1-\cos2F)F'A
-\frac{4}{r^2}(1-\cos2F)F'\frac{d^2B}{dr^2}
\notag \\
&-\frac{8}{r^2}\left[(1-\cos2F)\left(F''+\frac{3}{r}F'\right)
+\sin2F\,(F')^2\right]\drv{B}{r}
\notag \\
&-\frac{8}{r^3}\left[
(1-\cos 2F)\left(4F''+\frac{3}{r}F'-\frac{1}{r^2}\sin 2F\right)
+4\sin 2F\,(F')^2\right]B
\notag\\
&-\frac{2}{r^3}(1-\cos2F)F'\drv{C}{r}
-\frac{4}{r^3}\left[
(1-\cos 2F)\left(F''+\frac{2}{r}F'-\frac{1}{r^2}\sin 2F\right)
+\sin 2F\,(F')^2\right]C
\Biggr\} ,
\label{EQ1}
\\
\EQ_2&= -F'\drv{^2A}{r^2}
-2\left(F''+\frac{4}{r}F'\right)\drv{A}{r}
+\frac{2}{r}\left[-3 F''+\frac{1}{r}(\cos2F-3)F'+\frac{1}{r^2}\sin2F
\right]A
\notag \\
&+\left(F'-\frac{1}{2r}\sin2F\right)\drv{^2C}{r^2}
+\left[2F''+\frac{1}{r}(7-\cos2F)F'-\frac{3}{r^2}\sin2F
\right]\drv{C}{r}
\notag \\
&+\frac{2}{r}\left[3 F''+\frac{2}{r}(1-\cos2F)F'\right]C
\notag\\
&+\frac{1}{\fpi^2 e^2}\Biggl\{
\frac{2}{r^3}\left[(1-\cos 2F)\left(
F''+\frac{2}{r}F'-\frac{2}{r^2}\sin 2F\right)
+2\sin 2F\,(F')^2\right]
\notag\\
&-\frac{4}{r^2}(1-\cos2F)F'\drv{^2A}{r^2}
-\frac{4}{r^2}\left[2(1-\cos 2F)\left(F''+\frac{3}{r}F'\right)
+\sin 2F\,(F')^2\right]\drv{A}{r}
\notag \\
&-\frac{4}{r^3}\left[(1-\cos 2F)\left(8F''+\frac{1}{r}(1-2\cos 2F)F'
-\frac{2}{r^2}\sin 2F\right) +5\sin 2F\,(F')^2\right]A
\notag \\
&+\frac{1}{r^2}(1-\cos 2F)\left(4F'-\frac{1}{r}\sin 2F
\right)\drv{^2C}{r^2}
\notag\\
&+\frac{2}{r^2}\left[(1-\cos 2F)\left(4F''+\frac{1}{r}(7-2\cos 2F)F'
-\frac{2}{r^2}\sin 2F\right)+2\sin 2F\,(F')^2\right]\drv{C}{r}
\notag\\
&+\frac{4}{r^3}\left[(1-\cos 2F)\left(5F''-\frac{4}{r}(1+\cos 2F)F'
+\frac{2}{r^2}\sin 2F\right)+4\sin 2F\,(F')^2\right]C\Biggr\} ,
\label{EQ2}
\\
\EQ_3&=2\cos F\times\EQ_{34} ,
\label{EQ3}
\\
\EQ_4&=-2\sin F\times\EQ_{34} ,
\label{EQ4}
\end{align}
where $\EQ_{34}$ in \eqref{EQ3} and \eqref{EQ4} is given by
\begin{align}
\EQ_{34}&=\frac{1}{2r^3}\sin F +\frac{2}{r^2}\cos F\,F'A
-\frac{1}{2r}\sin F\drv{^2C}{r^2}
-\frac{1}{r}\!\left(\cos F\,F'+\frac{3}{r}\sin F\right)\!\drv{C}{r}
\notag \\
&-\frac{1}{r^2}\left(4\cos F\,F'+\frac{1}{r}\sin F\right)C
+\frac{1}{\fpi^2 e^2}\sin F\Biggl\{
\frac{2}{r^3}\left[(F')^2-\frac{1}{r^2}(1-\cos 2F)\right]
\notag \\
&+\frac{4}{r^2}(F')^2\drv{A}{r}
+\frac{4}{r^3}F'\left(3F'+\frac{2}{r}\sin 2F\right)A
\notag \\
&-\frac{1}{r^3}(1-\cos2F)\drv{^2C}{r^2}
-\frac{4}{r^2}\left[(F')^2+\frac{1}{r}\sin 2F\,F'
+\frac{1}{r^2}(1-\cos 2F)\right]\drv{C}{r}
\notag \\
&-\frac{4}{r^3}\left[
(F')^2+\frac{4}{r}\sin 2F\,F'-\frac{1}{r^2}(1-\cos 2F)
\right]C\Biggr\} .
\label{EQ34}
\end{align}
Next, the combination $\EQ_Y$ \eqref{defEQY} is given by
\begin{align}
\EQ_Y&=-3\,\EQ_1+\EQ_2-\EQ_3
\notag \\
&=\left(1+\frac{8}{\fpiesq}\frac{\sin^2F}{r^2}\right)
F'\drv{^2Y}{r^2}
\notag\\
&+\left\{2F''+\frac{8}{r}F'
+\frac{8}{\fpiesq}
\left[2\frac{\sin^2F}{r^2}\left(F''+\frac{3}{r}F'\right)
+\frac{\sin2F}{r^2}(F')^2\right]\right\}\drv{Y}{r}
\notag \\
&+\left\{\frac{6}{r}F''+\frac{14}{r^2}F'-\frac{2\sin 2F}{r^3}
+\frac{16}{\fpiesq}\!\left[\frac{\sin^2 F}{r^3}\left(
4 F''+\frac{3}{r}F'-\frac{\sin 2F}{r^2}\right)
+\frac{2\sin 2F}{r^3}(F')^2\right]\right\}Y
\notag \\
&-\frac{1}{2r^3}\sin2F + \frac{2}{\fpiesq}\left[
\frac{2\sin^2 F}{r^3}\left(F''+\frac{2}{r}F'-\frac{\sin 2F}{r^2}
\right) +\frac{\sin 2F}{r^3}(F')^2\right] .
\label{EQY}
\end{align}

By using \eqref{F_infty} for $F(r)$,
the differential equations $\EQ_2=\EQ_{34}=0$ for $(A,C)$ and
$\EQ_Y=0$ for $Y$ are approximated near the infinity $(r\to\infty)$
as follows:
\begin{align}
&\left(1+\frac{2}{s}+\frac{2}{s^2}\right)\drv{^2A}{s^2}
+2\left(-1+\frac{1}{s}+\frac{2}{s^2}+\frac{2}{s^3}
\right)\drv{A}{s}
-2\left(\frac{3}{s}+\frac{7}{s^2}+\frac{12}{s^3}+\frac{12}{s^4}
\right)A
\notag\\
&\qquad
-\left(1+\frac{3}{s}+\frac{3}{s^2}\right)\drv{^2C}{s^2}
+2\left(1-\frac{3}{s^2}-\frac{3}{s^3}\right)\drv{C}{s}
+6\left(\frac{1}{s}+\frac{3}{s^2}+\frac{6}{s^3}+\frac{6}{s^4}
\right)C=0 ,
\notag\\[3mm]
&-4\left(\frac{1}{s}+\frac{2}{s^2}+\frac{2}{s^3}\right)A
-\left(1+\frac{1}{s}\right)\drv{^2C}{s^2}
+2\left(1-\frac{1}{s}-\frac{1}{s^2}\right)\drv{C}{s}
+2\left(\frac{4}{s}+\frac{7}{s^2}+\frac{7}{s^3}\right)C
\notag\\
&\qquad
+\frac{1}{s^2}+\frac{1}{s^3}=0 ,
\notag\\
&\left(1+\frac{2}{s}+\frac{2}{s^2}\right)\drv{^2Y}{s^2}
+2\left(-1+\frac{1}{s}+\frac{2}{s^2}+\frac{2}{s^3}
\right)\drv{Y}{s}
-2\left(\frac{3}{s}+\frac{2}{s^2}+\frac{2}{s^3}+\frac{2}{s^4}
\right)Y+\frac{1}{s^3}+\frac{1}{s^4}=0 ,
\label{DE_ABC_infty}
\end{align}
where $s$ is the dimensionless variable $s=\mpi r$.
The general solution to \eqref{DE_ABC_infty} is given symbolically by
\eqref{ABCinfty} as the sum of a particular solution and six
independent modes. More concrete expression keeping terms up to
$1/s^5$ for the power parts is follows:
\begin{align}
\Pmatrix{A(r)\\ C(r)}
&=-\frac{1}{2s}\Pmatrix{1+3/s^3\\ 1-1/s+3/s^2+3/s^3}
+d_2\frac{1}{s^2}\Pmatrix{1-3/s^2\\ 1-2/s-4/s^2}
+d_3\frac{1}{s^3}\Pmatrix{1+2/s\\ 2+4/s}
\notag\\
&\quad
+\left\{
f_2\frac{1}{s^2}\Pmatrix{1-17/(3s^2)\\ 1-4/(3s)-2/(3s^2)}
+f_3\frac{1}{s^3}\Pmatrix{3-14/s\\ 2-8/s}
\right\}e^{2s} ,
\notag\\
Y(r)&=\frac{1}{2s^2}+d_3^Y\left(\frac{1}{s^3}-\frac{1}{s^4}\right)
+f_3^Y\left(\frac{1}{s^3}-\frac{3}{s^4}\right)e^{2s}
\qquad(r\to\infty).
\label{DE_ABC_infty_series}
\end{align}

\section{Derivation of  eq.\ (\ref{calJ}) for $\bm{\calJ}$}
\label{app:calc_calJ}

In this appendix, we outline the derivation of eq.\ \eqref{calJ} for
$\calJ$, the coefficient of the $\bmOmg^4$ term in the Lagrangian of
$R(t)$.
Here, as in Appendix \ref{app:deriv_FEOM_OSS}, $r$ denotes
the length of $\bmy$; $r=\abs{\bmy}$.
As explained in Sec.\ \ref{sec:RLagrangian}, calculation of
$\calJ$ is reduced via \eqref{calJ=(1/2)calJ_1} to that of $\calJ_1$,
the part of $\calJ$ linear in $(A,B,C)$.
It is convenient to use the expression of the Lagrangian $L$ as the
integration over $\bmy$ instead of $\bmx$:
\begin{equation}
L=\int\!d^3y\,J\,\calL\Bigr|_{U(\bmx,t)=\Ucl(\bmy)} ,
\label{L=intd^3y}
\end{equation}
where $J$ is the Jacobian and is given via \eqref{py/px} in terms of
$\yx{ab}$:
\begin{equation}
J=\abs{\det_{(a,i)}\left(\Drv{y_a}{x_i}\right)}^{-1}
=\Bigl|\det_{(a,b)}\left(\delta_{ab}+\yx{ab}\right)\Bigr|^{-1}
=1-\Tr\yx{}+O(\p_t^4) .
\end{equation}
Explicitly, we have
\begin{equation}
\Tr\yx{}=\yx{aa}=\left(5A-2C+r\drv{}{r}(A-C)\right)(\RiRd\bmy)^2
+r^2\left(5B+C+r\drv{B}{r}\right)\Tr(\RiRd)^2
\label{Tryx}
\end{equation}

Let us identify the part of $J\calL$ in \eqref{L=intd^3y} which is
{\em quartic} in $\bmOmg$ and {\em linear} in $(A,B,C)$ (we call such
part ``$\QLp$'' hereafter). We employ \eqref{LmubyLcl} for $L_\mu$
together with \eqref{py/pt} and \eqref{py/px} for $\p y_a/\p t$ and
$\p y_a/\p x_i$, respectively.
First, for the $\tr L_\mu^2$ part of $\calL$ \eqref{calL_Skyrme}, we
have
\begin{align}
J\tr L_0^2\Bigr|_{\QLp}&=
-\Tr\yx{}\,(\RiRd\bmy)_a(\RiRd\bmy)_b
\tr\bigl(\Lcl_a(\bmy)\Lcl_b(\bmy)\bigr) ,
\label{JtrL_0^2}
\\
J\tr L_i^2\Bigr|_{\QLp}&=0 .
\label{JtrL_i^2}
\end{align}
In obtaining \eqref{JtrL_0^2}, we have used the fact that the terms in
\eqref{py/pt} containing $\dtRiRd$ \eqref{dtRiRd} do not contribute
to the Lagrangian \eqref{L=intd^3y}, which we showed in Sec.\
\ref{sec:RLagrangian}.
This can of course be confirmed by explicit calculation.
For \eqref{JtrL_i^2}, we have used that the $O(\p_t^4)$ term
of $\yx{ab}$, which is not given in \eqref{yx}, is quadratic $(A,B,C)$
and hence cannot contribute to the $\QLp$.
Similarly, for the Skyrme-term
$\tr[L_\mu,L_\nu]^2=-2\tr[L_0,L_i]^2+\tr[L_i,L_j]^2$,
we have
\begin{align}
J\tr [L_0,L_i]^2\Bigr|_{\QLp}&=-\Bigl[\delta_{cd}\Tr\yx{}
-(\yx{cd}+\yx{dc})\Bigr](\RiRd\bmy)_a(\RiRd\bmy)_b
\tr\bigl([\Lcl_a,\Lcl_c][\Lcl_b,\Lcl_d]\bigr)(\bmy) ,
\label{Jtr[L_0,L_i]^2}
\\
J\tr[L_i,L_j]^2\Bigr|_{\QLp}&=0 .
\label{Jtr[L_i,L_j]^2}
\end{align}

Then, the $\QLp$ of the Lagrangian \eqref{L=intd^3y} is obtained by
using \eqref{JtrL_0^2}--\eqref{Jtr[L_i,L_j]^2}, \eqref{yx},
\eqref{Tryx} and
\begin{align}
\tr\bigl(\Lcl_a\Lcl_b\bigr)(\bmy)&=2(F')^2\why_a\why_b
+\frac{1-\cos 2F}{r^2}\left(\delta_{ab}-\why_a\why_b\right),
\\[3mm]
\tr\bigl([\Lcl_a,\Lcl_c][\Lcl_b,\Lcl_d]\bigr)(\bmy)
&=-2\frac{1-\cos 2F}{r^2}\,\delta_{ab}\left(2(F')^2\why_c\why_d
+\frac{1-\cos 2F}{r^2}\left(\delta_{cd}-\why_c\why_d\right)
\right)
\notag\\
&\quad
+2\left(\frac{1-\cos 2F}{r^2}\right)^2\delta_{ad}\delta_{bc}
+\left(\mbox{terms containing $\why_a$ and/or $\why_b$}\right) ,
\end{align}
and carrying out the solid-angle integration of $\bmy$.
Denoting by overline the solid-angle average,
\begin{equation}
\overline{\calO}=\frac{1}{4\pi}\int\!d\Omega_{\bmy}\,\calO ,
\end{equation}
the following formulas are of use:
\begin{align}
\overline{y_a y_b}&=\frac{r^2}{3}\delta_{ab} ,
\\
\overline{y_a y_b y_c y_d}&=\frac{r^4}{15}\left(\delta_{ab}\delta_{cd}
+\delta_{ac}\delta_{bd}+\delta_{ad}\delta_{bc}\right) .
\end{align}
In particular, we have
\begin{align}
\overline{(\RiRd\bmy)^2}&=-\frac{r^2}{3}\Tr(\RiRd)^2
=\frac23 r^2\bmOmg^2 ,
\label{olRiRdbmy^2}
\\
\overline{(\RiRd\bmy)^4}&=\frac{r^4}{15}\left[
\left(\Tr(\RiRd)^2\right)^2+2\Tr(\RiRd)^4\right]
=\frac{8}{15}r^4\bmOmg^4 ,
\end{align}
where we have used
\begin{align}
\Tr(\RiRd)^2&=-2\bmOmg^2 ,
\\
(\RiRd)^3&=\frac12\Tr(\RiRd)^2\,\RiRd .
\end{align}
By identifying the $\QLp$ of \eqref{L=intd^3y} with
$(1/4)\calJ_1\bmOmg^4$, we obtain $\calJ=(1/2)\calJ_1$ as given by
\eqref{calJ}.

\section{Equivalence of charges from $\bm{L(R,\dot{R})}$ and from the
Skyrme field theory}
\label{app:equiv_NoetherQ}

In this Appendix, we present a proof of the equivalence of the
conserved charges (especially, the isospin $\bm{I}$ and the
Hamiltonian $H$) obtained in two different ways; one is from the
Lagrangian $L(R,\dot{R})$ \eqref{L_R} of $R(t)$, and the other is by
the substitution of $U(\bmx,t)=\Ucl(\bmy)$ \eqref{OSS} into the
Noether charges from the Lagrangian density \eqref{calL_Skyrme} of the
Skyrme field theory.
We show that the two conserved charges agree with each other up to the
EOM of $R(t)$ and that in the Skyrme field theory; the latter also
vanishes upon using the EOM of $R(t)$ due to our principle.

\noindent
\underline{\em Isospin charge $I_a$}

First, let us consider the isospin charge $I_a$.
For this, recall that the Lagrangian of $R$ is related to the
Lagrangian density $\calL(U(\bmx,t))$ of the Skyrme model by
\begin{equation}
L(R,\dot{R})=\int\!d^3x\,\calL\bigr|_{U(\bmx,t)=\Ucl(\bmy)} .
\label{recall_L=intcalL}
\end{equation}
The isospin charge $I_a$ on the $R$-system side is obtained by
considering the following time-dependent transformation
$\dlmbd$ acting on $L$:
\begin{equation}
\dlmbd R(t)=R(t)\lambda(t) ,\quad
\dlmbd R^{-1}(t)=-\lambda(t)R^{-1}(t) ,
\label{TDT}
\end{equation}
where $\lambda_{ab}(t)$ is an infinitesimal anti-symmetric matrix.
We have
\begin{equation}
\dlmbd L(R,\dot{R})=\dot{\lambda}_a(t)I_a ,
\label{I_a-R}
\end{equation}
with $\lambda_a(t)=(1/2)\epsilon_{abc}\lambda_{bc}(t)$.

For relating the isospin charge of \eqref{I_a-R} to the one on the
field theory side, let us consider the response of
$\calL(U(\bmx,t)=\Ucl(\bmy))$ under the transformation
$\dlmbd$ of \eqref{TDT}.
Since we have $\dlmbd(R^{-1}\bmx)=-\lambda R^{-1}\bmx$ and
$\dlmbd(\RiRd)=\dot{\lambda}+\bigl[\RiRd,\lambda\bigr]$,
the transformation of $\bmy$ \eqref{bmy=} consists of two parts:
\begin{align}
\dlmbd\bmy=-\lambda\bmy + \dot{\lambda}_{ab}\bm{Z}_{ab}(\bmx,t) ,
\label{TYI}
\end{align}
where $\bm{Z}_{ab}$ is a three dimensional vector for each $(a,b)$:
\begin{equation}
(\bm{Z}_{ab})_c=\left[A(\RiRd\,\zbmy)_a\zy_b
-Br^2(\RiRd)_{ab}\right]\zy_c
+\frac12 Cr^2\left[
\delta_{ac}(\RiRd\,\zbmy)_b+(\RiRd)_{ca}\zy_b\right]
-(a\leftrightarrow b) ,
\end{equation}
with $\zbmy=R^{-1}\bmx$.
The first part $\dlmbd^{(0)}\bmy=-\lambda\bmy$ induces the standard
isospin transformation on $U(\bmx,t)=\Ucl(\bmy)$:
\begin{equation}
\dlmbd^{(0)}\Ucl(\bmy)=i\left[\Ucl(\bmy),
\lambda_a(t)\frac{\tau_a}{2}\right] ,
\label{SIT}
\end{equation}
and therefore we have
\begin{equation}
\dlmbd^{(0)}\calL\bigl(U(\bmx,t)=\Ucl(\bmy)\bigr)
=\dot\lambda_{a}J_{V,a}^{\mu=0}\Bigr|_{U(\bmx,t)=\Ucl(\bmy)} ,
\label{dlmbd^(0)calL=}
\end{equation}
where $J_{V,a}^\mu$ is the isovector current \eqref{JV} of the Skyrme
field theory. (If $\lambda_a$ in \eqref{SIT} depends on both $t$ and
$\bmx$, we have
$\dlmbd^{(0)}\calL=\p_\mu\lambda_a(\bmx,t)\,J_{V,a}^\mu$
instead of \eqref{dlmbd^(0)calL=}.)

The second part of the transformation \eqref{TYI},
$\dlmbd^{(1)}\bmy=\dot{\lambda}_{ab}\bm{Z}_{ab}(\bmx,t)$,
needs careful treatment.
Let us divide the Lagrangian density \eqref{calL_Skyrme} into two
parts,
$\calL=\calL_0(L_\mu) +(\fpi^2/8)\mpi^2\tr\left(U-\bm{1}_2\right)$,
and consider first the transformation of $\calL_0(L_\mu)$ consisting
only of $L_\mu=-i\Ucl(\bmy)(\p/\p x^\mu)\Ucl^\dagger(\bmy)$.
Using
\begin{equation}
\dlmbd^{(1)}\Ucl(\bmy)=\dot{\lambda}_{ab}(\bm{Z}_{ab})_c
\Drv{}{y_c}\Ucl(\bmy) ,
\label{dlmbd^(1)Ucl}
\end{equation}
we have
\begin{equation}
\dlmbd^{(1)}L_\mu
=\Drv{\eta}{x^\mu}+i\left[L_\mu(\bmx,t),\eta\right] ,
\label{dlmbd^(1)L_mu}
\end{equation}
with $\eta$ defined by
\begin{equation}
\eta=\dot{\lambda}_{ab}(\bm{Z}_{ab})_c\Lcl_c(\bmy).
\label{eta}
\end{equation}
This gives
\begin{equation}
\dlmbd^{(1)}\calL_0(L_\mu)=\tr\!\left(
\Drv{\calL_0}{L_\mu}\,\dlmbd^{(1)}L_\mu\right)
=\tr\!\left(\Drv{\calL_0}{L_\mu}\Drv{\eta}{x^\mu}\right)
\label{dlmbd^(1)calL_0}
\end{equation}
with
\begin{equation}
\Drv{\calL_0}{L_\mu}=-\frac{\fpi^2}{8}\left(L^\mu
-\frac{1}{\fpiesq}\bigl[L_\nu,[L^\mu,L^\nu]\bigr]\right) .
\end{equation}
In $\dlmbd^{(1)}\!\int\!d^3x\,\calL_0$, we carry out the
integration-by-parts for the $\mu=1,2,3$ parts of
\eqref{dlmbd^(1)calL_0}. For the $\mu=0$ part, notice that
$\p/\p t$ acting on $(\bm{Z}_{ab})_c\Lcl_c(\bmy)$ in \eqref{eta},
which consists of $\zbmy$ and $\RiRd$, can be expressed as
\begin{equation}
\Drv{}{t}=-(\dot{R}R^{-1}\bmx)_i\Drv{}{x_i}+(\mbox{$R$-EOM terms}) ,
\label{(p/pt)=}
\end{equation}
where ($R$-EOM terms) denote terms which vanish upon use of the EOM
\eqref{REOM} of $R$ (c.f., eq.\ \eqref{EDbmy=0}).
Since \eqref{(p/pt)=} holds also against $\p\calL_0/\p L_0$ and since
we have $(\dot{R}R^{-1})_{ii}=0$, space-integration-by-parts for all
$\mu$ in \eqref{dlmbd^(1)calL_0} is effectively allowed to give
\begin{equation}
\dlmbd^{(1)}\!\int\!d^3x\,\calL_0=(\mbox{$\ddot{\lambda}$-term})
-\int\!d^3x\,\tr\!\left(\eta\,\Drv{}{x^\mu}\Drv{\calL_0}{L_\mu}\right)
+(\mbox{$R$-EOM terms}).
\label{dlmbd^(1)intcalL_0}
\end{equation}
The $\ddot{\lambda}$-term is another contribution from the $\mu=0$
term of \eqref{dlmbd^(1)calL_0}; $\p/\p t$ acting on
$\dot{\lambda}_{ab}$ in $\eta$ \eqref{eta}. However, this
$\ddot{\lambda}$-term in fact vanishes. This is because, after
carrying out the solid-angle integration of $\zbmy$, the possible
quantities multiplying anti-symmetric matrix $\ddot{\lambda}_{ab}$ are
all symmetric ones,
$\delta_{ab}\Tr(\RiRd)^2$ and $\bigl((\RiRd)^2\bigr)_{ab}$,
in our present approximation of keeping at most four time-derivatives
in the Lagrangian (see the argument given in
Sec.\ \ref{sec:RLagrangian} for excluding $(d/dt)(\RiRd)$ from the
Lagrangian of $R$ ).

For the $\dlmbd^{(1)}$ transformation of the remaining pion mass term
in $\calL$ \eqref{calL_Skyrme}, we have, using \eqref{dlmbd^(1)Ucl},
\begin{equation}
\dlmbd^{(1)}\tr\left(U-\bm{1}_2\right)
=\dot{\lambda}_{ab}(\bm{Z}_{ab})_c\tr\Drv{\Ucl(\bmy)}{y_c}
=-i\tr\left(\eta\,\Ucl(\bmy)\right) .
\end{equation}
{}From this and \eqref{dlmbd^(1)intcalL_0}, we have the following for
the whole $\calL$:
\begin{equation}
\dlmbd^{(1)}\!\int\!d^3x\,\calL=-\int\!d^3x\,\tr\left\{
\eta\left[\Drv{}{x^\mu}\Drv{\calL_0}{L_\mu}
+i\frac{\fpi^2}{8}\mpi^2\left(U-\frac12\tr U\right)\right]
\right\}_{U=\Ucl(\bmy)}
+(\mbox{$R$-EOM terms}).
\label{dlmbd^(1)intcalL}
\end{equation}
Note that the quantity inside the square bracket multiplying $\eta$ in
the first term on the RHS is nothing but the EOM \eqref{FEOM} of the
Skyrme field theory. Due to our EOM principle, this term also vanishes
upon use of the EOM of $R$.

We have calculated the $\dlmbd$ transformation of the both hand sides
of \eqref{recall_L=intcalL}; the transformation of the LHS is
\eqref{I_a-R}, while that of the RHS is the sum of (the
space-integration of) \eqref{dlmbd^(0)calL=} and
\eqref{dlmbd^(1)intcalL}. An important point is that the whole of
\eqref{dlmbd^(1)intcalL} vanishes when the EOM \eqref{REOM} of $R$
holds. Therefore, comparing the coefficients of $\dot{\lambda}_a(t)$,
we find that the isospin charge $I_a$ obtained from the Lagrangian
\eqref{L_R} is equal to that in the Skyrme field theory up to the EOM
of $R$:
\begin{equation}
I_a=\int\!d^3x\,J_{V,a}^{\mu=0}(\bmx,t)\Bigr|_{U(\bmx,t)=\Ucl(\bmy)}
+(\mbox{$R$-EOM terms}) .
\end{equation}

For the spin $J_i$, we consider the infinitesimal left transformation
on $R(t)$ instead of the right one \eqref{TDT}. Then the equivalence
of $J_i$ between the $R$ system and the Skyrme field theory can be
shown quite similarly.

\noindent
\underline{\em Hamiltonian $H$}

Next, we show that the Hamiltonian \eqref{H} of the system of $R$ is
equal to the Hamiltonian of the Skyrme field theory with the
substitution \eqref{OSS} up to the EOM of $R$. The proof is quite
similar to the above one for the isospin, and we will describe only
the points of the proof.

By considering the time translation with infinitesimal time-dependent
parameter
$\veps(t)$,
\begin{equation}
\dtt R(t)=\veps(t)\dot{R}(t) ,
\label{dtt}
\end{equation}
the Hamiltonian \eqref{H} of the $R$ system is obtained as
\begin{equation}
\dtt L=\dot{\veps}\,H +\p_t(\veps L).
\end{equation}
To relate this $H$ to the Hamiltonian of the Skyrme field theory, we
consider the action of the same transformation \eqref{dtt} on the
Lagrangian density on the RHS of \eqref{recall_L=intcalL}.
We have
\begin{equation}
\dtt\bmy(\bmx,t)=\veps(t)\p_t\bmy(\bmx,t)
+\dot{\veps}(t)\bm{z}(\bmx,t) ,
\end{equation}
with
\begin{equation}
\bm{z}=2\left[A(\RiRd\zbmy)^2 +Br^2\Tr(\RiRd)^2
+Cr^2(\RiRd)^2\right]\zbmy .
\end{equation}
The first part of the transformation,
$\dtt^{(0)}\bmy=\veps\p_t\bmy$, gives simply
\begin{equation}
\dtt^{(0)}\!\int\!d^3x\,\calL\bigl(U(\bmx,t)=\Ucl(\bmy)\bigr)
=\dot{\veps} H_\text{Skyrme}\Bigr|_{U(\bmx,t)=\Ucl(\bmy)}
+\p_t(\veps L) ,
\end{equation}
with $H_\text{Skyrme}$ being the Hamiltonian of the Skyrme field
theory. For the second part of the transformation,
$\dtt^{(1)}\bmy=\dot{\veps}\bm{z}$, the same equation as
\eqref{dlmbd^(1)L_mu} holds by replacing $\eta$ \eqref{eta} with
$\dot{\veps}z_a\Lcl_a(\bmy)$. Then, the same argument by using
\eqref{(p/pt)=} applies also to $\dtt^{(1)}$, leading again to
\eqref{dlmbd^(1)intcalL}.
In this case, the absence of the $\ddot{\veps}$ term is owing to that
the possible quantities multiplying $\ddot{\veps}$ are $\Tr(\RiRd)^3$
and $\Tr(\RiRd)^2\Tr(\RiRd)$, which however all vanish identically.
In this way, we obtain the equivalence of $H$ and
$H_\text{Skyrme}|_{U(\bmx,t)=\Ucl(\bmy)}$.

The equivalence of the Hamiltonians obtained by two ways, which holds
owing to the EOM principle, provides us with another and simple
derivation of the relation \eqref{Virial}.
First, the Hamiltonian $H$ corresponding to the Lagrangian
\eqref{L=1+Omg^2+Omg^4} is given by
\begin{equation}
H=\Mcl+\frac12\left(\calI+\Delta\calI\right)\bmOmg^2
+\frac34\left(\calJ_1+\calJ_2\right)\bmOmg^4 .
\end{equation}
On the other hand, the Hamiltonian of the Skyrme field theory is
obtained from the Lagrangian by inverting the sign of its ``potential
term'' $V$ \eqref{Vpart}:
\begin{equation}
H_\text{Skyrme}\Bigr|_{U(\bmx,t)=\Ucl(\bmy)}
=T+V
=\Mcl+\frac12\left(\calI-\Delta\calI\right)\bmOmg^2
+\frac14\left(\calJ_1-\calJ_2\right)\bmOmg^4 .
\end{equation}
Comparing the above two, we reobtain \eqref{Virial}.

\section{Analytic expressions of the static properties of nucleons}
\label{app:AE}

In this appendix, we outline the derivation of the analytic
expressions of the various static properties of the nucleons.
Throughout this appendix, $r$ denotes the length of $\bmy$,
$r=\abs{\bmy}$, except $\VEV{r^2}(=\VEV{\bmx^2})$ denoting the charge
radii of various kinds.

\noindent
\underline{\em Isoscalar charge radius}

The isoscalar mean square charge radius $\VEV{r^2}_{I=0}$ is given
in terms of the baryon number density $J_B^0(\bmx,t)$ and a nucleon
state $\ket{N}$ by
\begin{equation}
\VEV{r^2}_{I=0}=\bra{N}\int\! d^3x\,\bmx^2 J_B^0(\bmx,t)\ket{N} .
\label{def_VEVr^2_I=0}
\end{equation}
{}From \eqref{JB} and \eqref{LmubyLcl}, we have
\begin{equation}
J_B^0(\bmx,t)=-\frac{i}{24\pi^2}\epsilon_{ijk}
\tr\left(L_i(\bmx,t)L_j(\bmx,t)L_k(\bmx,t)\right)
=\abs{\det_{(a,i)}\!\left(\Drv{y_a}{x_i}\right)}
\JBclz(\abs{\bmy}) ,
\label{J_B^0(bmx,t)=}
\end{equation}
where $\JBclz$ is the baryon number density of the static solution
\eqref{hedgehog}:
\begin{equation}
\JBclz(r)=-\frac{1}{4\pi^2}\frac{1-\cos 2F}{r^2}\,\drv{F(r)}{r} .
\end{equation}
Owing to the presence of the Jacobian in \eqref{J_B^0(bmx,t)=}, it is
convenient to switch to the $\bmy$-integration to evaluate
\eqref{def_VEVr^2_I=0} as follows:
\begin{align}
\VEV{r^2}_{I=0}&=\bra{N}\int\! d^3y\,\bmx^2 \JBclz(\abs{\bmy})\ket{N}
\notag\\
&=4\pi\int_0^\infty\! dr\,r^4\bra{N}\Bigl[
1-2\left(A-C\right)\overline{(\RiRd\bmy)^2}
-2B r^2\Tr(\RiRd)^2\Bigr]\ket{N}\JBclz(r)
\notag\\
&=\frac{4\pi}{\fpiesq}\int_0^\infty\!d\rdl\,\rdl^4\left(
1+\frac43 e^4\bmOmgdl_N^2\rdl^2 Y\right)\JBclz(\rdl)
\label{ISCR}
\end{align}
where we have used the relation
\begin{equation}
\bmx^2=\left[1-2\left(A-C\right)(\RiRd\bmy)^2
-2B r^2\Tr(\RiRd)^2\right]r^2
\qquad(r=\abs{\bmy}),
\label{bmx^2=}
\end{equation}
valid to $O(\p_t^2)$, and \eqref{olRiRdbmy^2} for the angle average
over $\bmy$. The final expression is given in terms of the
dimensionless $\rdl=e\fpi r$, $\bmOmgdl_N^2$ \eqref{valueOmegas}
and the function $Y(r)$ \eqref{defY}.

\noindent
\underline{\em Isovector charge radius}

To obtain the expression of the isovector charge radius, we first need
that of the isospin charge density $J_{V,a}^0$ of \eqref{JV} with
$J_{L/R}^\mu$ defined by \eqref{eq:J_LJ_R}.
Corresponding to \eqref{LmubyLcl} for $L_\mu$,
$R_\mu=-iU^\dagger\p_\mu U$ for our spinning Skyrmion field
\eqref{OSS} is given by
\begin{equation}
R_0(\bmx,t)=\Drv{y_a(\bmx,t)}{t}\Rcl_a(\bmy),
\qquad
R_i(\bmx,t)=\Drv{y_a(\bmx,t)}{x_i}\Rcl_a(\bmy) ,
\label{RmubyRcl}
\end{equation}
with
\begin{align}
\Rcl_a(\bmy)&=-i\Ucl(\bmy)^\dagger\Drv{}{y_a}\Ucl(\bmy)
\notag\\
&=\frac{1}{2r}\Bigl\{\sin 2F\delta_{ab}
+\left(2rF'-\sin 2F\right)\why_a\why_b
+\left(1-\cos 2F\right)\epsilon_{abc}\why_c\Bigr\}\tau_b
\notag\\
&=-\Lcl_a(-\bmy) .
\label{Rcl_a=}
\end{align}
Note that $\Ucl(\bmy)^\dagger=\Ucl(-\bmy)$.
Since we are considering the on-shell $R(t)$ with $\dtRiRd=0$,
\eqref{py/pt} is reduced to $\p y_a(\bmx,t)/\p t=-(\RiRd\bmy)_a$
in the present calculation. This leads to
\begin{equation}
L_0+R_0=\frac{1}{r}\left(1-\cos 2F\right)
\epsilon_{abc}(\RiRd\bmy)_a\why_b\tau_c .
\label{L_0+R_0}
\end{equation}
As for the part in $J_V^0$ from the Skyrme term, we have
\begin{equation}
\bigl[L_i,[L_0,L_i]\bigr]+(L_\mu\to R_\mu)
=-\left(\delta_{ac}+\yx{ac}+\yx{ca}\right)
(\RiRd\bmy)_b T_{abc}
+(\bmy\to -\bmy) ,
\label{[L_i[L_0,L_i]]+(L->R)}
\end{equation}
with $\yx{ac}+\yx{ca}$ and $T_{abc}$ given by \eqref{yx_ab+yx_ba}
and \eqref{T_abc}, respectively. From \eqref{L_0+R_0} and
\eqref{[L_i[L_0,L_i]]+(L->R)}, we obtain
\begin{align}
&16 e^2J_V^0(\bmx,t)
=\fpiesq(L_0+R_0)-\Bigl(\bigl[L_i,[L_0,L_i]\bigr]
+\bigl[R_i,[R_0,R_i]\bigr]\Bigr)
\notag\\
&=2\,\frac{1-\cos 2F}{r}\Biggl\{\frac12\fpiesq
+2(F')^2+\frac{1-\cos 2F}{r^2}
\notag\\
&\quad
+\left[4\left(3(A-C)+r\drv{}{r}(A-C)\right)(F')^2
+2(A+C)\frac{1-\cos 2F}{r^2}\right](\RiRd\bmy)^2
\notag\\
&\quad
+\left[4\left(3B+r\drv{B}{r}\right)r^2(F')^2
+(2B+C)(1-\cos 2F)\right]\Tr(\RiRd)^2\Biggr\}
\epsilon_{abc}(\RiRd\bmy)_a\why_b\tau_c .
\label{J_V^0(bmx,t)=}
\end{align}
Using this, we get the following expression for the isovector mean
square charge radius of the nucleon:
\begin{align}
&\VEV{r^2}_{I=1}=\frac{
\bra{N}\int\!d^3x\,\bmx^2 J_{V,a=3}^0(\bmx,t)\ket{N}}{
\bra{N}I_3\ket{N}}
\notag\\
&=\frac{4\pi}{3\sqrt{3}}\,\frac{\bigl|\bmOmgdl_N\bigr|}{\fpiesq}
\int_0^\infty\!d\rdl\,\rdl^4\sin^2 F\Biggl\{
1+4(F')^2+4\frac{\sin^2 F}{\rdl^2}
\notag\\
&\quad
+\frac25 e^4\bmOmgdl_N^2\rdl^2\biggl[\rdl Z'+7Z-C
+4(F')^2\left(-\rdl Z'+Z-C\right)
+4\frac{\sin^2 F}{\rdl^2}\left(\rdl Z'+5Z+2C\right)\biggr]
\Biggr\},
\label{IVCR}
\end{align}
with prime denoting $d/d\rdl$ and $Z$ defined by \eqref{Z}.
In deriving \eqref{IVCR}, it is convenient to switch to the
$\bmy$-integration by using $d^3x=\left(1-\Tr\yx{}\right)d^3y$
with $\Tr\yx{}$ given by \eqref{Tryx}, and use \eqref{bmx^2=} for
$\bmx^2$ and the following formulas for the angle averaging:
\begin{align}
\epsilon_{abc}\,\overline{(\RiRd\bmy)_b\why_c}
&=\frac13 r\epsilon_{abc}(\RiRd)_{bc}=\frac23 r\Omega_a ,
\notag\\
\epsilon_{abc}\,\overline{(\RiRd\bmy)^2(\RiRd\bmy)_b\why_c}
&=-\frac{2}{15}r^3\Tr(\RiRd)^2\epsilon_{abc}(\RiRd)_{bc}
=\frac{8}{15}r^3\bmOmg^2\Omega_a
\label{ol_Omega}
\end{align}
In rewriting $\Omega_3$ in the numerator of \eqref{IVCR} in terms of
$I_3$, we have used \eqref{I^2=()^2Omg^2} for the nucleon, namely,
\begin{equation}
\left(\calI+\calJ\bmOmg_N^2\right)^{-1}
=\frac{2}{\sqrt{3}}\abs{\bmOmg_N} .
\label{(I+JOmg^2)^-1=}
\end{equation}

Here, we comment on another derivation of \eqref{calJ} for $\calJ$.
In Sec.\ \ref{sec:detfpi}, the isospin charge $I_a$ \eqref{I_a=}
was obtained from the Lagrangian \eqref{L_R} of $R(t)$.
As explained in Appendix \ref{app:equiv_NoetherQ}, the same $I_a$
should also be obtained by integrating \eqref{J_V^0(bmx,t)=} over
$\bmx$. By the latter way, we directly obtain \eqref{calJ} for $\calJ$
(and also \eqref{calI} for $\calI$) without using the relation
\eqref{calJ=(1/2)calJ_1}.

\noindent
\underline{\em Isoscalar magnetic moment and magnetic charge radius}

Let us consider the isoscalar magnetic moment defined by
\begin{equation}
\bm{\mu}_{I=0}=\frac12\int\!d^3x\,\bmx\times\bm{J}_B(\bmx,t) .
\label{bmu_I=0}
\end{equation}
The space component of the baryon number current \eqref{JB} is given
by
\begin{equation}
J_B^i(\bmx,t)=\frac{i}{8\pi^2}\epsilon_{ijk}
\tr(L_j L_k L_0)
=\left(1+\Tr\yx{}\right)R_{ia}
\left(\delta_{ab}-\yx{ab}\right)(\RiRd\bmy)_b\JBclz(\abs{\bmy}) ,
\end{equation}
where we have used \eqref{LmubyLcl}, \eqref{py/px}, \eqref{py/pt} with
$\dtRiRd=0$, and
\begin{equation}
\tr\left(\Lcl_a\Lcl_b\Lcl_c\right)=4\pi^2i\epsilon_{abc}\JBclz .
\end{equation}
Again, \eqref{bmu_I=0} is most easily evaluated by switching to
$\bmy$-integration and expressing $\bmx$ in terms of $\bmy$ via
\begin{equation}
\zbmy=R^{-1}\bmx
=\left[1-A(\RiRd\bmy)^2-B r^2\Tr(\RiRd)^2\right]\bmy
-C r^2(\RiRd)^2\bmy .
\end{equation}
Then, the isoscalar $g$-factor $g_{I=0}$ of the nucleon defined by
\begin{equation}
\bm{\mu}_{I=0}=\frac{g_{I=0}}{2 M_N}\bm{J} ,
\end{equation}
is found to be given by
\begin{equation}
g_{I=0}=\frac{16\pi}{3\sqrt{3}}\frac{e M_N}{\fpi}
\,\bigl|\bmOmgdl_N\bigr|
\int_0^\infty\!d\rdl\,\rdl^4 \JBclz(\rdl)
\left[1+\frac25e^4\bmOmgdl_N^2
\rdl^2\left(2Z+C\right)\right] .
\label{g_I=0}
\end{equation}
The isoscalar mean square magnetic radius $\VEV{r^2}_{M,I=0}$, which
is defined as \eqref{bmu_I=0} with extra weight $\bmx^2$ divided by
\eqref{bmu_I=0} itself, is given by
\begin{equation}
\fpiesq\VEV{r^2}_{M,I=0}=\frac{
\int_0^\infty\!d\rdl\,\rdl^6\JBclz(\rho)\left[1+
\frac25 e^4\bmOmgdl_N^2\rdl^2\left(4Z+C\right)\right]}{
\int_0^\infty\!d\rdl\,\rdl^4\JBclz(\rho)\left[1+
\frac25 e^4\bmOmgdl_N^2\rdl^2\left(2Z+C\right)\right]} .
\label{ISMCR}
\end{equation}

\noindent
\underline{\em Isovector magnetic moment and magnetic charge radius}

Calculation of the isovector magnetic moment,
\begin{equation}
\bm{\mu}_{I=1}=\int\!d^3x\,\bmx\times\bm{J}_{V,a=3}(\bmx,t) ,
\label{bmmu_I=1}
\end{equation}
is much more involved. Relaxing the $a=3$ restriction in
\eqref{bmmu_I=1} to a generic $a$, the $i$-th component
$(\bm{\mu}_{I=1})_i$ is found to have three kinds of dependence on
$(i,a)$;
$R_{ia}$, $\bigl(R(\RiRd)^2\bigr)_{ia}$ and $\omega_i\Omega_a$.
When we consider only the nucleon matrix elements,
we have the following representation \cite{AN}:
\begin{equation}
J_i=\frac12\sigma_i\otimes 1,\quad I_a=1\otimes\frac12\tau_a,\quad
R_{ia}=-\frac13\sigma_i\otimes\tau_a .
\end{equation}
Then, owing to \eqref{(I+JOmg^2)^-1=}, we have
\begin{equation}
\omega_i\Omega_a=\left(\calI+\calJ\bmOmg_N^2\right)^{-2}
\frac14\,\sigma_i\otimes\tau_a
=\frac13\bmOmg_N^2\,\sigma_i\otimes\tau_a ,
\end{equation}
and consequently
\begin{equation}
\bigl(R(\RiRd)^2\bigr)_{ia}=-\omega_i\Omega_a-\bmOmg^2R_{ia}
=0 ,
\label{R(RiRd)^2=0}
\end{equation}
both for the nucleon states.
Using these facts, the isovector $g$-factor $g_{I=1}$ of the nucleon
defined by
\begin{equation}
\left(\bm{\mu}_{I=1}\right)_i=\frac{g_{I=1}}{2 M_N}
\frac{\bm{\sigma}}{2}\otimes\tau_3 ,
\end{equation}
is calculated to be given by
\begin{equation}
g_{I=1}=\frac{8\pi}{9}\frac{M_N}{e^3\fpi}\,\wt{g}_{I=1} ,
\label{g_I=1}
\end{equation}
with
\begin{align}
\wt{g}_{I=1}&=\int_0^\infty
\!d\rdl\,\rdl^2\sin^2 F\Biggl\{1+4(F')^2+4\frac{\sin^2 F}{\rdl^2}
\notag\\
&\quad
+\frac25 e^4\bmOmgdl_N^2\rdl^2\left[
\rdl Z'+5Z-C-4(F')^2\left(\rdl Z'+Z+C\right)
+4\frac{\sin^2 F}{\rdl^2}\left(\rdl Z'+3Z+2C\right)\right]
\Biggr\} .
\label{wtg_I=1}
\end{align}
Similarly, the isovector mean square magnetic radius
$\VEV{r^2}_{M,I=1}$ is given by
\begin{align}
&\fpiesq\VEV{r^2}_{M,I=1}=\frac{1}{\wt{g}_{I=1}}
\int_0^\infty\!d\rdl\,\rdl^4\sin^2 F\,\Biggl\{
1+4(F')^2+4\frac{\sin^2 F}{\rdl^2}
\notag\\
&\qquad
+\frac25 e^4\bmOmgdl_N^2\rdl^2\left[\rdl Z'+7Z-C
+4(F')^2\left(-\rdl Z'+Z-C\right)
+4\frac{\sin^2 F}{\rdl^2}\left(\rdl Z'+5Z+2C\right)\right]\Biggr\} .
\label{IVMCR}
\end{align}

Comparing \eqref{wtg_I=1} for $\wt{g}_{I=1}$ with \eqref{calI} for
$\calI$ and \eqref{calJ} for $\calJ$, we find the following simple
relationship:
\begin{equation}
\wt{g}_{I=1}=\frac{3}{2\pi}
\left(\calIdl+e^4\bmOmgdl_N^2\calJdl\right)
=\frac{3\sqrt{3}}{4\pi}\bigl|\bmOmgdl_N\bigr|^{-1} ,
\label{simpleR}
\end{equation}
where the last equality is due to \eqref{(I+JOmg^2)^-1=}.
This implies that $g_{I=1}$ has a much simpler expression:
\begin{equation}
g_{I=1}=\frac{2}{\sqrt{3}}\frac{M_N}{e^3\fpi}
\bigl|\bmOmgdl_N\bigr|^{-1} ,
\label{g_I=1_simpler}
\end{equation}
and that the magnetic radius $\VEV{r^2}_{M,I=1}$ is in fact equal to
the electric one $\VEV{r^2}_{I=1}$ \eqref{IVCR}:
\begin{equation}
\VEV{r^2}_{M,I=1}=\VEV{r^2}_{I=1} .
\label{IVMCR=IVCR}
\end{equation}
The relations \eqref{g_I=1_simpler} and \eqref{IVMCR=IVCR} are also
valid in the rigid body approximation and should have a simple origin.

\noindent
\underline{\em Axial vector coupling}

Let us consider the axial vector current $J^\mu_{A,a}$ \eqref{JA}.
The axial vector coupling constant $g_A=g_A(0)$ is obtained by
identifying the nucleon matrix elements of
\begin{equation}
\int\!d^3x\,J_{A,a}^i(\bmx,t) ,
\label{intJ_A}
\end{equation}
with the $\bm{q}=\bm{p}'-\bm{p}\to 0$ limit of
\begin{equation}
\bra{N'(\bm{p}')}J_{A,a}^i(0)\ket{N(\bm{p})}=\frac12\,g_A(\bm{q}^2)
\bra{N'}\sigma_i\otimes\tau_a\ket{N} ,
\end{equation}
valid for nonzero $\mpi$.
The evaluation of \eqref{intJ_A} is also very complicated.
Using the relation \eqref{R(RiRd)^2=0}, we obtain
\begin{equation}
g_A=\frac{2\pi}{9 e^2}\left(\wt{g}_A^{(1)}+\wt{g}_A^{(3)}\right) ,
\label{g_A}
\end{equation}
where $\wt{g}_A^{(1)}$ and $\wt{g}_A^{(3)}$, which are contributions
from the part of $J^i_{A,a}$ linear and cubic in $L_\mu$ or $R_\mu$,
respectively, are given by
\begin{align}
\wt{g}_A^{(1)}&=-\int_0^\infty\!d\rdl\,\rdl\left\{
\rdl F'+\sin 2F+\frac15 e^4\bmOmgdl_N^2\rdl^2\left[
2\rdl F'+\sin 2F\left(4+\rdl\drv{}{\rdl}\right)\right]
\left(2Z+C\right)\right\} ,
\label{wtg_A^(1)}
\\
\wt{g}_A^{(3)}&=4\int_0^\infty\!d\rdl\,\rdl\,\sin 2F\Biggl\{
-(F')^2-F'\frac{\tan F}{\rdl}-\frac{\sin^2 F}{\rdl^2}
\notag\\
&\qquad
+\frac15 e^4\bmOmgdl_N^2\biggl[-2\rdl^2(F')^2\left(
-\rdl Z'+5A-10B-4C\right)
\notag\\
&\qquad
+\rdl F'\tan F\left(\rdl C'+2A+2C+1\right)
+\sin^2 F\left(
4-\Bigl(2+\rdl\drv{}{\rdl}\Bigr)(2Z+C)\right)\biggr]
\Biggr\} .
\label{wtg_A^(3)}
\end{align}

\noindent
\underline{\em Isoscalar quadrupole moment}

The isoscalar quadrupole moment operator $Q^{I=0}_{ij}$,
\begin{equation}
Q^{I=0}_{ij}=\int\!d^3 x\left(x_ix_j-\frac13\delta_{ij}\bmx^2\right)
J_B^0(\bmx,t) ,
\end{equation}
is a measure of the non-spherical deformation due to the spinning
motion. $Q^{I=0}_{ij}$ vanishes identically in the rigid body
approximation where the baryon number density is spherically
symmetric \cite{AN2}.
In the present case with relativistic correction, it becomes
non-trivial:
\begin{equation}
\fpiesq Q^{I=0}_{ij}=\frac{8\pi e^4}{15}\int_0^\infty\!
d\rdl\,\rdl^6\left(2A-5C\right)\JBclz(\rdl)
\frac{\bmOmgdl^2}{\bm{J}^2}\left(
\frac12\left\{J_i,J_j\right\}-\frac13\delta_{ij}\bm{J}^2\right) .
\label{ISQM}
\end{equation}
However, the nucleon expectation value of the operator
$\frac12\left\{J_i,J_j\right\}-\frac13\delta_{ij}\bm{J}^2$
and hence of \eqref{ISQM} are equal to zero.

\end{document}